\documentclass[twocolumn]{aastex701}

\newboolean{showrevisions}
\setboolean{showrevisions}{true} % 切换 true/false
\newcommand{\revise}[1]{%
  \ifthenelse{\boolean{showrevisions}}{\textcolor{red}{#1}}{#1}%
}

\newcommand{\mathd}{\mathrm{d}}
\shorttitle{TransFit-MAG}
\shortauthors{Li et al.}
\graphicspath{{./}{figures/}}
\usepackage{amsmath}
\usepackage{threeparttable}
\usepackage{booktabs}
\usepackage{changepage} 

\graphicspath{{./}{figures/}}
\begin{document}

\title{ \texttt{TransFit-MAG}: Self-Consistent Modeling of Magnetar-Powered Transients from Shock Breakout to Spin-Down Heating}

\author[0009-0004-9719-272X]{Jing-Yao Li}
\affiliation{Institute of Astrophysics, Central China Normal University, Wuhan 430079, China; \url{yuyw@ccnu.edu.cn;liuld@ccnu.edu.cn}}
\affiliation{Laboratory for Compact Object Astrophysics and Astronomical Technology, Central China Normal University, Wuhan 430079, China}
\affiliation{Education Research and Application Center, National Astronomical Data Center, Wuhan 430079, China}
\email{lijy@mails.ccnu.edu.cn}

\author[0000-0002-8708-0597]{Liang-Duan Liu}
\affiliation{Institute of Astrophysics, Central China Normal University, Wuhan 430079, China; \url{yuyw@ccnu.edu.cn;liuld@ccnu.edu.cn}}
\affiliation{Laboratory for Compact Object Astrophysics and Astronomical Technology, Central China Normal University, Wuhan 430079, China}
\affiliation{Education Research and Application Center, National Astronomical Data Center, Wuhan 430079, China}
\email{liuld@ccnu.edu.cn}

\author[0000-0002-1067-1911]{Yun-Wei Yu}
\affiliation{Institute of Astrophysics, Central China Normal University, Wuhan 430079, China; \url{yuyw@ccnu.edu.cn;liuld@ccnu.edu.cn}}
\affiliation{Laboratory for Compact Object Astrophysics and Astronomical Technology, Central China Normal University, Wuhan 430079, China}
\affiliation{Education Research and Application Center, National Astronomical Data Center, Wuhan 430079, China}
\email{yuyw@ccnu.edu.cn}

\author[0000-0001-8744-3813]{Guang-Lei Wu}
\affiliation{Institute of Astrophysics, Central China Normal University, Wuhan 430079, China; \url{yuyw@ccnu.edu.cn;liuld@ccnu.edu.cn}}
\affiliation{Laboratory for Compact Object Astrophysics and Astronomical Technology, Central China Normal University, Wuhan 430079, China}
\affiliation{Education Research and Application Center, National Astronomical Data Center, Wuhan 430079, China}
\email{wuguanglei@mails.ccnu.edu.cn}

\author[0009-0000-2423-6825]{Yu-Hao Zhang}
\affiliation{Institute of Astrophysics, Central China Normal University, Wuhan 430079, China; \url{yuyw@ccnu.edu.cn;liuld@ccnu.edu.cn}}
\affiliation{Laboratory for Compact Object Astrophysics and Astronomical Technology, Central China Normal University, Wuhan 430079, China}
\affiliation{Education Research and Application Center, National Astronomical Data Center, Wuhan 430079, China}
\email{zhang-yh@mails.ccnu.edu.cn}

\begin{abstract}
Magnetar engines are widely invoked to power luminous optical transients, but
their early emission depends on the coupled evolution of engine injection,
shock heating, adiabatic cooling, and radiative diffusion. We present
\texttt{TransFit-MAG}, a time-dependent radiative-diffusion framework for
magnetar-powered transients. The model couples the \texttt{TransFit} diffusion
solver to the dynamics of a magnetar-inflated pulsar wind nebula (PWN) and its
forward shock propagating through homologously expanding ejecta, 
%. Unlike Arnett-like semi-analytic models, \texttt{TransFit-MAG} follows 
calculating the internal
radiation-energy distribution, photospheric evolution, shock-heating location,
and emergent luminosity self-consistently. 
%, without assuming a fixed diffusion time or imposing an artificial delay between energy injection and photon escape. We show that the light-curve morphology is controlled by the competition between shock-heated radiation escaping from the outer ejecta and delayed diffusion of magnetar/PWN energy deposited at deeper layers. A clear double peak requires a sufficiently early and luminous shock-breakout component relative to the rising magnetar-powered diffusion component. 
%Depending on the ejecta mass, opacity, magnetar energy, and spin-down timescale, 
For different parameter values, the model
naturally produces well-separated double peaks, partially merged peaks, or
single broad peaks. %, which provide a natural explanation for the observed diverse morphologies of the light curves of engine-powered transients. 
These results suggest that early bumps and broad single peaks in engine-powered transients may be understood within a unified engine--shock--diffusion framework, in which the observed diversity reflects the coupled evolution of central-engine power, shock propagation, and radiative transport through expanding ejecta.
As an illustrative application, we fit the multiband
optical light curves of the double-peaked SLSN-I LSQ14bdq. 
%The early bump is reproduced as shock-heated emission, while the broad main peak is powered by delayed diffusion of magnetar/PWN energy. 
\end{abstract}


\section{Introduction}
\label{sec:introduction}

High-cadence time-domain surveys have revealed a diverse population of luminous and rapidly evolving optical transients \citep{bellm2018,ivezic2019}, including superluminous supernovae \citep[SLSNe;][]{gal2019}, fast blue optical transients \citep[FBOTs;][]{drout2014,pursiainen2018}, and supernovae associated with gamma-ray bursts (GRBs) \citep{galama1998,hjorth2003,stanek2003,woosley2006} or extragalactic fast X-ray transients \citep{sun2025,li2025}. 
%The supernovae associated with long-duration GRBs are broad-lined Type Ic explosions, as exemplified by GRB~980425/SN~1998bw and GRB~030329/SN~2003dh \citep{galama1998,hjorth2003,stanek2003,woosley2006}. Recent extragalactic fast X-ray transient--supernova associations also point to Type Ic-BL explosions, suggesting a related engine-driven massive-star origin \citep{sun2025,li2025}. 
These events occupy a broad region of luminosity--timescale phase space \citep{Inserra2019} and often show rapid color evolution, early-time excess emission, or multiwavelength counterparts \citep{margutti2019,ho2019}. 
%Although the radioactive decay chain $^{56}{\rm Ni}\rightarrow^{56}{\rm Co}\rightarrow^{56}{\rm Fe}$ powers the light curves of normal supernovae \citep{arnett1982}, 
The high luminosities, rapid evolution, and short diffusion timescales of many of these transients challenge a purely radioactive interpretation due to the decay chain $^{56}{\rm Ni}\rightarrow^{56}{\rm Co}\rightarrow^{56}{\rm Fe}$. 
Additional power from a central engine, such as a rapidly rotating magnetar \citep{kasen2010,woosley2010,yu2013,yu2015,zhang2022} or accretion onto a compact object\citep{Piro2011,Dexter2013,Moriya2019,Lin2021}, is therefore often invoked.

Magnetar spin-down provides a natural energy reservoir for powering luminous optical transients \citep{yu2017,liu2022}. 
In the standard picture, the spin-down luminosity is injected into the expanding ejecta, thermalized, and released through radiative diffusion. 
However, the early emission depends not only on the instantaneous spin-down power, but also on how this energy is deposited and transported. 
A magnetar wind can inflate a hot pulsar-wind nebula (PWN), which acts as a piston and drives a shock into the homologously expanding ejecta \citep{chevalier2005,kasen2016,chen2016,suzuki2021}. 
The shock converts part of the relative kinetic energy into thermal radiation, while the remaining magnetar/PWN energy is stored in the optically thick ejecta and released after a diffusion delay. 
Recent work has further emphasized that, if the wind bubble reaches the steep outer ejecta before photon diffusion becomes efficient, hydrodynamic instabilities may lead to a blowout of the nascent bubble and produce a fast early UV/optical peak \citep{Chen2026}. 
The observed light curve is therefore shaped by the coupled evolution of engine injection, PWN-driven shock heating, adiabatic expansion, and radiative diffusion.

This coupling is particularly relevant for hydrogen-poor SLSNe with double-peaked rising light curves. 
Several SLSNe-I show an early precursor bump before the broader main peak \citep{Leloudas2012,Nicholl2015,Smith2016}. 
A systematic search by \citet{Nicholl2016} found plausible early bumps in 8 of 14 SLSNe with sufficiently early observations, suggesting that such features may be common among SLSNe-I. 
However, the larger Dark Energy Survey sample shows that early bumps are not ubiquitous \citep{Angus2019}. 
The presence or absence of an early bump therefore provides a useful diagnostic of the energy-deposition depth, optical-depth evolution, and diffusion efficiency of the ejecta.

Despite these observational motivations, magnetar-powered light curves are
still commonly modeled with semi-analytic diffusion prescriptions, including
Arnett-type solutions \citep{arnett1980,arnett1982} and their extensions to
magnetar heating \citep{kasen2010,chatzopoulos2012,nicholl2017}. These
models are useful for estimating global parameters, but 
%they usually describe photon escape with a fixed effective diffusion time or a prescribed response function. As a result, 
they do not explicitly follow the spatially dependent
energy deposition and transport in the ejecta. This limitation becomes
particularly important when the magnetar inflates a PWN
and drives a shock through the expanding ejecta. In this case, the shock
radius, the optical depth above the shocked layer, and the local
shock-heating rate all evolve with time. An artificial delay between magnetar
injection and radiative escape is sometimes introduced to reproduce
double-peaked light curves \citep{kasen2016,liu2021}, but a more
self-consistent treatment should instead follow the time-dependent
internal-energy distribution and its diffusion through the ejecta
\citep{pinto2000a,pinto2000b,blinnikov2006,morozova2015}, ideally guided by
radiation-hydrodynamic calculations \citep{dessart2012,kasen2016,chen2016}.

To address this problem, we previously developed \texttt{TransFit}, a
time-dependent radiative-diffusion framework for homologously expanding ejecta
\citep{Liu2025}. Rather than adopting an integral diffusion solution,
\texttt{TransFit} directly solves the diffusion equation and self-consistently
calculates the internal-energy profile, photospheric evolution, and emergent
luminosity. This framework has also been extended to interaction-powered
transients by coupling the diffusion solver to ejecta--CSM shock dynamics and
time-dependent shock heating \citep{zhang2026}. In this work, we extend \texttt{TransFit} to magnetar-powered transients. The
resulting model, \texttt{TransFit-MAG}, couples magnetar spin-down injection,
PWN-driven shock dynamics, local shock heating, and time-dependent radiative
diffusion in expanding ejecta, %. It follows the thermal evolution and light-curve morphology without imposing a fixed diffusion time or an
%artificial delay between energy injection and photon escape. This framework therefore 
providing a physically motivated tool for interpreting early bumps,
shoulders, and broad peaks in engine-powered transients.
%, and can be applied to the analysis of diverse optical transients discovered by large time-domain survey programs.

The paper is organized as follows. 
In Section~\ref{sec:Model}, we describe the physical framework of 
\texttt{TransFit-MAG}, including the homologously expanding ejecta structure, 
magnetar spin-down input, PWN-driven shock dynamics, and the coupled 
time-dependent radiative-diffusion treatment. 
In Section~\ref{sec:morphology_timescales}, we investigate how the competition 
between shock-heated radiation and delayed magnetar/PWN diffusion shapes the 
bolometric light curve, and we formulate the physical conditions for 
well-separated double peaks, partially merged peaks, and single broad peaks. 
In Section~\ref{sec:lsq14bdq_fit}, we apply the model to the double-peaked 
SLSN-I LSQ14bdq and compare the inferred parameters with previous studies. 
Finally, in Section~\ref{sec:discussion_conclusion}, we summarize our main 
results, discuss their physical implications, and outline the limitations of 
the present model.

\section{Physical Framework}
\label{sec:Model}

\begin{figure}[htbp!]
    \centering
    \includegraphics[width=1.1\columnwidth, trim=40 0 40 0, clip]{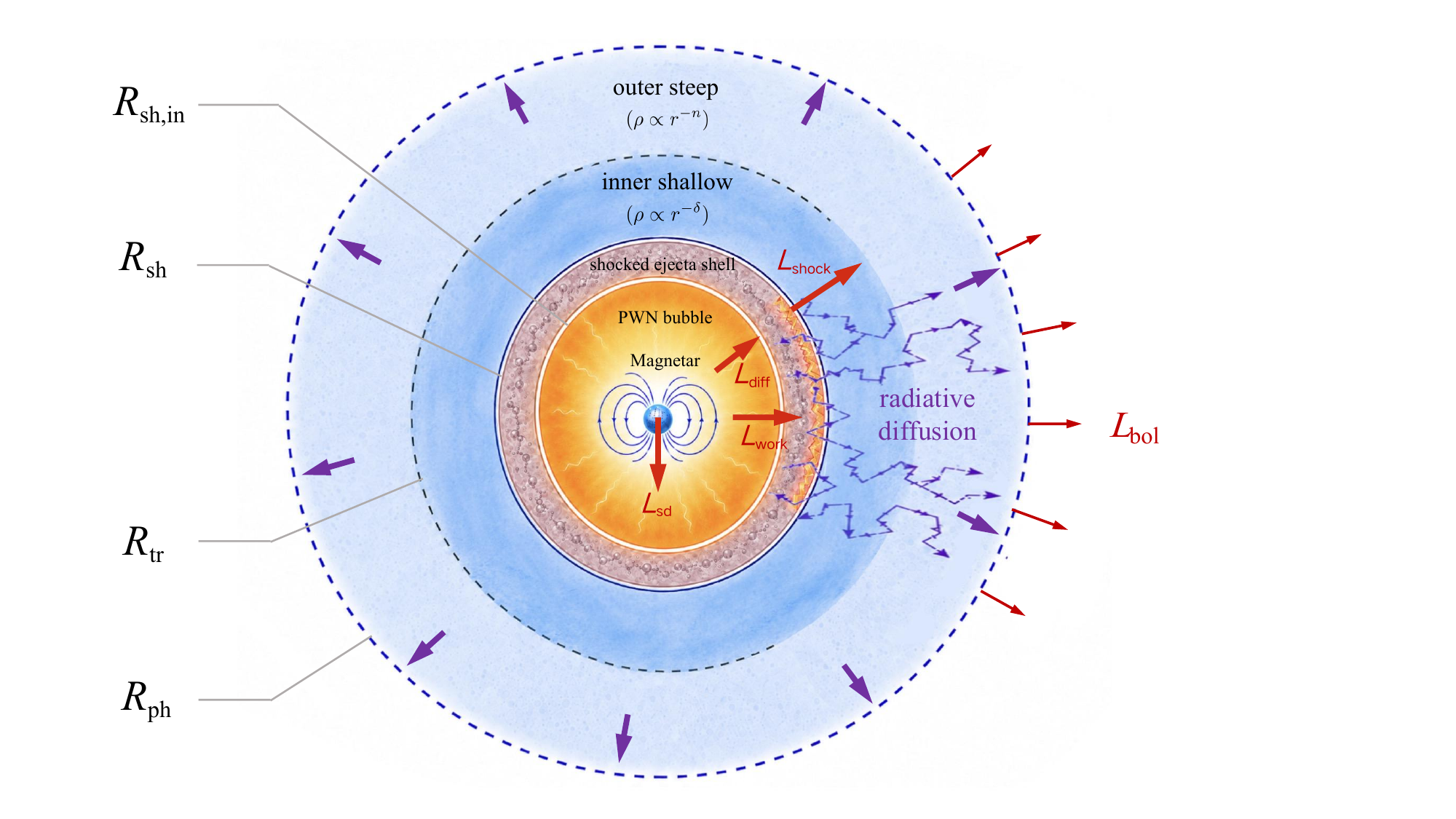}
    \caption{
    Schematic of the coupled PWN--shock--diffusion framework in \texttt{TransFit-MAG}. 
    The magnetar spin-down power $L_{\rm sd}$ inflates a radiation-dominated PWN bubble, which drives a forward shock through the homologously expanding ejecta and sweeps material into a shocked shell. 
    The bubble performs mechanical work $L_{\rm work}$ on the shell, while shock dissipation produces local heating $L_{\rm shock}$ near $R_{\rm sh}$. 
    The deposited radiation diffuses through the optically thick ejecta toward the photosphere $R_{\rm ph}$, beyond which photons free-stream and contribute to the observed bolometric luminosity $L_{\rm bol}$.
    }
    \label{fig:schematic}
\end{figure}

As an extension of the \texttt{TransFit} framework, \texttt{TransFit-MAG} models transients powered by the spin-down of a newly born magnetar embedded in expanding supernova ejecta. 
The magnetar deposits its rotational energy into the inner ejecta, % through magnetic dipole spin-down, 
forming a hot, radiation-dominated PWN,  
%Because the ejecta are initially optically thick, the injected energy cannot escape directly. It is instead stored as internal radiation energy, partly degraded by adiabatic expansion, and gradually released through radiative diffusion. The observed light curve therefore depends on the coupled evolution of central-engine injection, PWN-driven shock heating, and the time-dependent diffusion of radiation through the expanding ejecta.
as illustrated in Figure~\ref{fig:schematic}. % illustrates the physical picture implemented in \texttt{TransFit-MAG}. 
%The magnetar spin-down luminosity inflates the PWN bubble, which 
The PWN acts as a piston and drives a forward shock into the homologously expanding ejecta. 
Part of the PWN energy is converted into mechanical work on the swept-up shell, while the relative kinetic energy across the shock is dissipated locally as shock heating. 
The resulting radiation field is then redistributed within the optically thick ejecta and transported outward by diffusion until it reaches the photosphere, beyond which photons free-stream to form the observed emission.
This treatment differs from semi-analytic models that 
%approximate photon escape with a fixed diffusion timescale or 
assumes a prescribed global response function\citep{kasen2016,liu2021,Chen2026}. 
%Such approximations do not explicitly follow where energy is deposited, how it is stored and degraded by adiabatic expansion, or how the declining optical depth modifies photon escape. In \texttt{TransFit-MAG}, the PWN input, shock heating, internal-energy evolution, and photospheric boundary condition are evolved self-consistently. The effective delay between energy injection and photon escape is therefore an outcome of the coupled dynamical--diffusion calculation rather than an externally imposed parameter.

\subsection{Initial Conditions}
\label{sec:initial_conditions}

We specify the initial ejecta and engine properties adopted in \texttt{TransFit-MAG}. 
For core-collapse supernovae (CCSNe), the ejecta expand rapidly after shock breakout. 
When the ejecta radius has increased to several times the progenitor radius, the flow approaches homologous expansion, with $v=r/t$. 
We therefore start the calculation at $t_{\rm in}$, defined as the epoch at which the ejecta are assumed to have reached this homologous stage. 
The ejecta velocity is limited to the range
$v_{\rm min}\leq v\leq v_{\rm max}$.

The ejecta density is written as
\begin{equation}
\rho_{\rm ej}(r,t)
=
\rho_0
\left(\frac{t}{t_{\rm in}}\right)^{-3}
\eta_{\rm ej}(x),
\label{eq:rho_ej_profile}
\end{equation}
where $\rho_0$ is the density normalization at $t_{\rm in}$. 
The factor $(t/t_{\rm in})^{-3}$ describes the density dilution caused by homologous expansion. 
The dimensionless velocity coordinate is defined as
\begin{equation}
x \equiv \frac{v}{v_{\rm max}}
      = \frac{r}{v_{\rm max}t}.
\label{eq:x_ej_def}
\end{equation}

The function $\eta_{\rm ej}(x)$ gives the dimensionless density profile in velocity space. 
Motivated by hydrodynamic models of CCSN ejecta, we adopt a continuous broken power-law form \citep{chevalier1989,matzner1999},
\begin{equation}
\eta_{\rm ej}(x)=
\begin{cases}
\left(x/x_{\rm tr}\right)^{-\delta}, & x_{\min} \leq x < x_{\rm tr} ,\\
\left(x/x_{\rm tr}\right)^{-n},      & x_{\rm tr}\leq x \leq 1 .
\end{cases}
\label{eq:eta_ej}
\end{equation}
Here $x_{\min}=v_{\rm min}/v_{\rm max}$ and $x_{\rm tr}=v_{\rm tr}/v_{\rm max}$. 
The transition velocity $v_{\rm tr}$ separates the shallow inner ejecta from the steep outer envelope. 
In homologous expansion, this velocity corresponds to the transition radius $R_{\rm tr}(t)=v_{\rm tr}t$.
We adopt $\delta\simeq 1$ and $n\simeq 10$, typical values for compact progenitor explosions. 
The normalization $\rho_0$ and the transition velocity $v_{\rm tr}$ are determined by requiring the density profile to reproduce the adopted ejecta mass $M_{\rm ej}$ and the initial supernova kinetic energy $E_{\rm SN}$. 
Here $E_{\rm SN}$ denotes the kinetic energy carried by the homologously expanding ejecta at $t_{\rm in}$, before any subsequent energy injection from the central magnetar engine. 
The residual internal energy of the freely expanding ejecta at $t_{\rm in}$ is assumed to be subdominant, because it has been substantially reduced by adiabatic expansion and is rapidly overtaken by the subsequent shock-heating input.

The central engine is powered by the spin-down of a newborn magnetar. 
In the magnetic dipole approximation, the spin-down luminosity is
\begin{equation}
L_{\rm sd}(t)
=
\frac{E_{\rm m}}{t_{\rm sd}}
\left(1+\frac{t}{t_{\rm sd}}\right)^{-2},
\label{eq:Lsd}
\end{equation}
where $E_{\rm m}$ is the initial rotational-energy reservoir and $t_{\rm sd}$ is the characteristic spin-down timescale \citep{dai1998,kasen2010}. 
The magnetar energy $E_{\rm m}$ is therefore distinct from $E_{\rm SN}$: the former represents the energy reservoir available for central-engine injection, whereas the latter specifies the initial kinetic energy of the homologously expanding ejecta. 
The injected spin-down power is deposited into the inner ejecta, inflating a hot, radiation-dominated bubble that drives a shock into the surrounding expanding material.

\subsection{PWN-driven Shock Dynamics}
\label{sec:shock_dynamics}

The spin-down power of the central magnetar inflates a hot, high-pressure PWN inside the expanding ejecta. 
The PWN acts as a piston and drives a forward shock into the overlying ejecta \citep{chevalier1992,chevalier2005}. 
As the shock propagates outward, it sweeps up ejecta material into a compressed shell and converts part of the relative kinetic energy into thermal radiation.

Because the shocked layer is expected to remain narrow compared with the global expansion scale, and because we are primarily interested in the integrated momentum and energy exchange rather than the detailed internal structure of the shocked region, we adopt a thin-shell approximation.
We note that multidimensional effects may become important when the PWN-driven shell approaches the steep outer ejecta. 
In particular, hydrodynamic instabilities can fragment the shell and lead to a blowout of the nascent wind bubble \citep{Chen2026}. 
These effects are not explicitly modeled here; instead, we use the thin-shell dynamics as an effective description of the angle-averaged momentum and energy exchange.

Within this approximation, the shocked ejecta are characterized by the forward-shock radius $R_{\rm sh}$, velocity 
$v_{\rm sh}=\mathd R_{\rm sh}/\mathd t$, and swept-up mass $M_{\rm sh}$. 
The density of the unshocked ejecta immediately ahead of the shock is 
$\rho_{\rm sh}\equiv \rho_{\rm ej}(R_{\rm sh},t)$. 
Because the upstream ejecta expand homologously, their velocity at the shock position is $R_{\rm sh}/t$. 
We therefore define the shock velocity relative to the upstream ejecta as
$\Delta v_{\rm sh}\equiv v_{\rm sh}-R_{\rm sh}/t$. 
Only this relative velocity enters the mass flux, ram pressure, and shock-heating terms.
The shell evolution follows from mass conservation, radial momentum balance, and the energy budget of the PWN bubble. 
These equations describe the growth of the swept-up shell, its acceleration by the bubble pressure against the upstream ram pressure, and the competition between magnetar input, mechanical work, and radiative diffusion:
\begin{align}
\frac{\mathd M_{\rm sh}}{\mathd t}
&=
4\pi R_{\rm sh}^{2}\rho_{\rm sh}\Delta v_{\rm sh},
\label{eq:dMshdt}
\\
\frac{\mathd v_{\rm sh}}{\mathd t}
&=
\frac{4\pi R_{\rm sh}^{2}}{M_{\rm sh}}
\left(
P_{\rm b}
-
\rho_{\rm sh}\Delta v_{\rm sh}^{2}
\right),
\label{eq:dvshdt}
\\
\frac{\mathd U_{\rm b}}{\mathd t}
&=
L_{\rm sd}
-
L_{\rm work}
-
L_{\rm diff}.
\label{eq:dUbdt}
\end{align}
Here $P_{\rm b}$ and $U_{\rm b}$ denote the pressure and internal energy of the radiation-dominated PWN bubble. 
Using $U_{\rm b}=3P_{\rm b}V_{\rm b}$ with $V_{\rm b}=4\pi R_{\rm sh}^{3}/3$, we obtain
\begin{equation}
U_{\rm b}
=
4\pi R_{\rm sh}^{3}P_{\rm b}.
\label{eq:Ub_def}
\end{equation}

The mechanical power $L_{\rm work}$ represents the net energy transferred from the PWN bubble to the shocked region. 
It is not simply the time derivative of the shell kinetic energy, because the swept-up ejecta already carry bulk kinetic energy before crossing the shock. 
We therefore write
\begin{equation}
L_{\rm work}
=
\frac{\mathd E_{\rm k,sh}}{\mathd t}
-
\frac{\mathd E_{\rm k,ej}}{\mathd t}
+
L_{\rm shock},
\label{eq:Lwork}
\end{equation}
where $E_{\rm k,sh}$ is the kinetic energy of the shocked shell,$E_{\rm k,sh}
=
\frac{1}{2}M_{\rm sh}v_{\rm sh}^{2}$,

and $\mathd E_{\rm k,ej}/\mathd t$ is the rate at which the pre-existing kinetic energy of the homologously expanding ejecta is swept into the shell,
\begin{equation}
\frac{\mathd E_{\rm k,ej}}{\mathd t}
=
2\pi R_{\rm sh}^{2}\rho_{\rm sh}
\left(
\frac{R_{\rm sh}}{t}
\right)^{2}
\Delta v_{\rm sh}.
\label{eq:dEkejdt}
\end{equation}

The shock-heating luminosity is the dissipation rate of the relative kinetic energy across the shock \citep{kasen2016,li2016},
\begin{equation}
L_{\rm shock}
=
2\pi R_{\rm sh}^{2}\rho_{\rm sh}\Delta v_{\rm sh}^{3}.
\label{eq:Lshock}
\end{equation}

The radiation transmitted from the PWN bubble into the shocked ejecta is evolved by radiative diffusion. 
In the diffusion approximation, the luminosity through a spherical surface at radius $r$ is
\begin{equation}
L(r,t)
=
-4\pi r^{2}
\frac{c}{3\kappa\rho}
\frac{\partial u}{\partial r},
\label{eq:diff_luminosity}
\end{equation}
where $u(r,t)$ is the radiation energy density, $\kappa$ is the opacity, and $\rho$ is the local mass density in the diffusion region. 

At the inner boundary of the shocked ejecta, the diffusive luminosity supplied by the PWN is limited by the radiation-pressure contrast across the PWN--shell interface. 
We write
\begin{equation}
L_{\rm diff}
=
3\pi R_{\rm sh,in}^{2}c
\left(
P_{\rm b}-P_{\rm sh,in}
\right),
\label{eq:L_diff}
\end{equation}
where $P_{\rm b}$ is the PWN pressure and
\begin{equation}
P_{\rm sh,in}
=
\frac{u(R_{\rm sh,in},t)}{3}
\end{equation}
is the radiation pressure at the inner edge of the shocked ejecta. 
Equation~(\ref{eq:L_diff}) is obtained by multiplying the one-sided diffusive flux driven by the pressure contrast,
$F_{\rm diff}\simeq 3c(P_{\rm b}-P_{\rm sh,in})/4$, by the interface area $4\pi R_{\rm sh,in}^{2}$. 
Thus, $L_{\rm diff}$ measures the radiative power transported from the PWN bubble into the shocked ejecta, and is coupled to the evolving radiation field at the inner boundary.

In the thin-shell approximation, the inner radius of the shocked shell is related to the forward-shock radius by
\begin{equation}
R_{\rm sh,in}
=
(1-k_{\rm sh})R_{\rm sh},
\end{equation}
where $k_{\rm sh}\equiv \Delta R_{\rm sh}/R_{\rm sh}$ is the fractional shell thickness. 
For a compressed shell, we adopt
\begin{equation}
k_{\rm sh}
=
\frac{1}{(3-\delta)\chi_{\rm sh}},
\end{equation}
with $\chi_{\rm sh}=7$ for the shock compression ratio.

% \begin{equation}
% L_{\rm diff}=L_{\rm rad,sh}(t)
% =
% -4\pi R_{\rm sh}^{2}
% \frac{c}{3\kappa\rho_{\rm sh}}
% \left.
% \frac{\partial u}{\partial r}
% \right|_{r=R_{\rm sh}} .
% \label{eq:Lrad_sh}
% \end{equation}
% This quantity measures the radiation flux transported away from the shock-heated region by diffusion. 
% It is therefore not necessarily equal to the instantaneous shock-heating power $L_{\rm shock}$, because part of the dissipated energy can remain stored as internal radiation energy or be lost to adiabatic expansion.

\begin{figure}[htbp!]
    \centering
    \includegraphics[scale = 0.45]{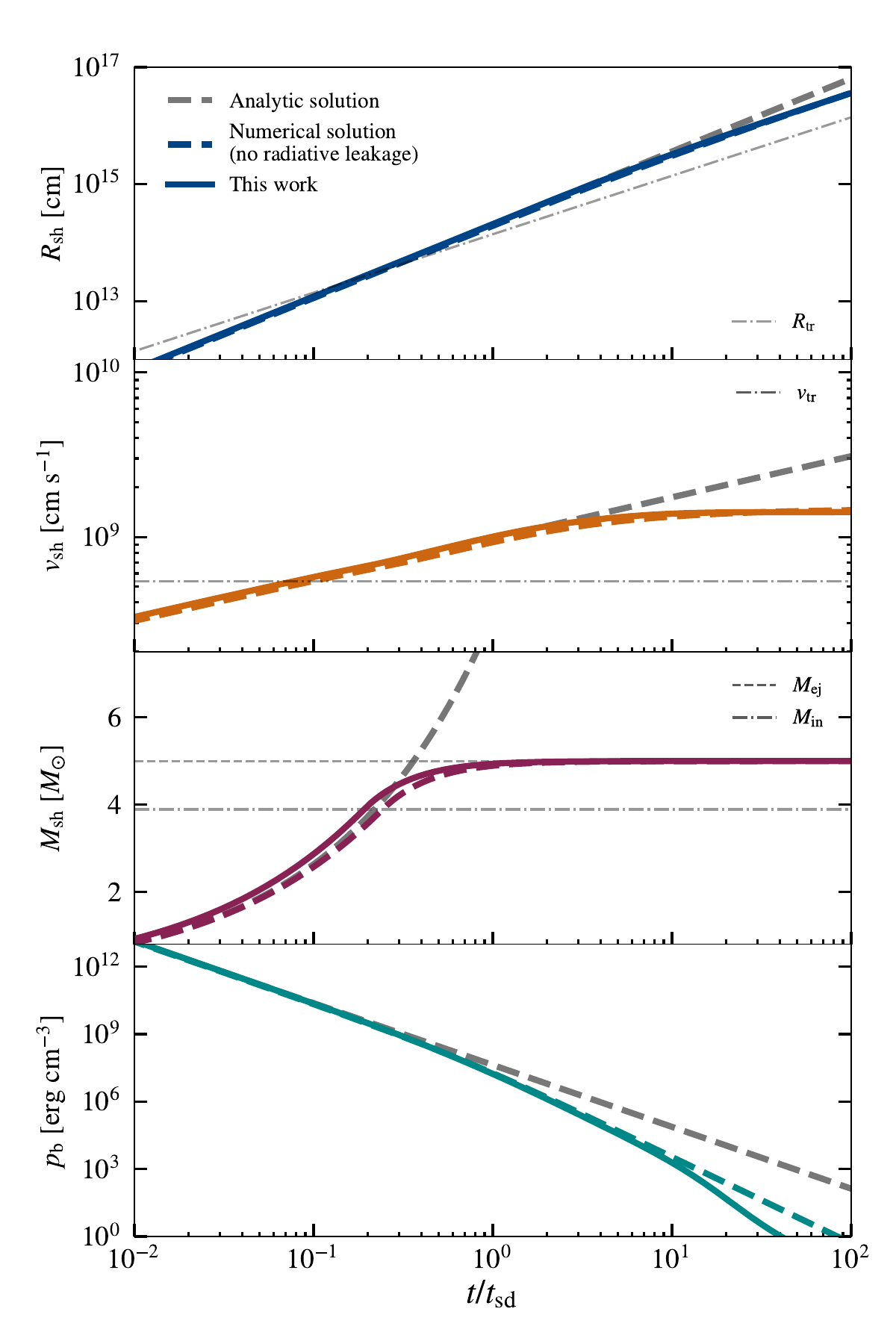}
  \caption{
Comparison of the PWN-driven shock evolution from the analytic self-similar solution of \citet{chevalier2005}, a numerical solution without radiative leakage, and the full dynamical--diffusion calculation developed in this work. 
From top to bottom, the panels show $R_{\rm sh}$, $v_{\rm sh}$, $M_{\rm sh}$, and $P_{\rm b}$ as functions of $t/t_{\rm sd}$. 
The dot-dashed lines mark $R_{\rm tr}$ and $v_{\rm tr}$, and the horizontal lines in the mass panel indicate $M_{\rm in}$ and $M_{\rm ej}$. 
The calculations adopt $M_{\rm ej}=5\,M_\odot$, $E_{\rm SN}=10^{51}\,{\rm erg}$, $E_{\rm m}=10^{52}\,{\rm erg}$, and $t_{\rm sd}=3\,{\rm days}$.
}
\label{fig:shock_dynamics}
\end{figure}

At early times, the shock remains in the shallow inner ejecta, the magnetar spin-down luminosity is nearly constant ($t\ll t_{\rm sd}$), and the ejecta are still optically thick. 
Radiative leakage from the PWN--shocked-ejecta system is therefore dynamically negligible, and the energy-driven self-similar solution approximately applies. 
For an inner ejecta density profile $\rho_{\rm ej}\propto t^{\delta-3}r^{-\delta}$, this gives $R_{\rm sh}\propto t^{(6-\delta)/(5-\delta)}$ \citep{chevalier2005}.

After the shock enters the steep outer ejecta, the swept-up mass grows more slowly and approaches the finite mass available above the density break. 
The dynamics then approach the limiting case of a pressure-driven shell with an approximately fixed mass. 
If the injected power remains nearly constant and radiative losses are still subdominant, $E_{\rm k,sh}\sim L_{\rm sd}t$, yielding $v_{\rm sh}\propto t^{1/2}$ and $R_{\rm sh}\propto t^{3/2}$ \citep{kasen2016}. 
In the full calculation, deviations from this simple scaling arise because $L_{\rm sd}$ decreases with time and radiative diffusion removes energy from the PWN--shocked-ejecta system.

If radiative leakage is neglected, the above dynamical equations form a closed set of ordinary differential equations and can be solved numerically \citep{chevalier2005,kasen2016,liu2021}. 
However, as the shock propagates outward and the ejecta expand, the optical depth decreases and radiative diffusion begins to remove energy from the bubble and shocked ejecta. 
In this regime, the shock dynamics must be coupled to the time-dependent diffusion equation. 
We therefore solve the full dynamical--diffusion system, in which the radiation field is evolved self-consistently and the resulting diffusive loss feeds back on the PWN pressure and shell acceleration.

Figure~\ref{fig:shock_dynamics} compares the PWN-driven shock evolution obtained from three approaches: the analytic self-similar solution of \citet{chevalier2005}, a numerical solution of the dynamical equations without radiative leakage \citep{kasen2016,liu2021}, and the full dynamical--diffusion calculation developed in this work.

The three solutions agree well while the shock remains in the shallow inner ejecta. 
The agreement breaks down when the shock approaches the transition radius $R_{\rm tr}$, or equivalently when the homologous ejecta velocity at the shock position, $R_{\rm sh}/t$, becomes comparable to $v_{\rm tr}$. 
The analytic solution assumes a single power-law ejecta density profile and therefore extrapolates the inner-ejecta self-similar behavior into the outer layers. 
In contrast, the numerical calculations include the broken power-law ejecta structure. 
Once the shock enters the steep outer envelope, the mass-loading rate and pressure-driven acceleration change, the swept-up mass approaches the available ejecta mass, and the shock velocity deviates from the inner-ejecta self-similar scaling.

This departure is already present in the no-leakage numerical solution, indicating that it is mainly caused by the density break rather than by radiative losses. 
At later times, the full dynamical--diffusion calculation further deviates from the no-leakage solution because radiative diffusion removes part of the magnetar-injected energy from the PWN--shocked-ejecta system. 
This leakage reduces the energy retained in the bubble, accelerates the decline of $P_{\rm b}$, and weakens the late-time driving of the shell.

\subsection{Radiative Diffusion and Energy Deposition}
\label{sec:radiative_diffusion}

To compute the emergent light curve, we follow the time-dependent thermal evolution of the ejecta under the combined effects of shock heating, PWN energy injection, adiabatic expansion, and radiative diffusion. 
Unlike semi-analytic models that prescribe the heating as a spatially integrated luminosity, our calculation evolves the radiation energy density $u(r,t)$ directly.

Assuming a radiation-dominated gas, the energy equation is \citep{Liu2025}
\begin{equation}
\frac{1}{\rho}
\frac{\partial u}{\partial t}
-
\frac{4u}{3\rho^{2}}
\frac{\partial \rho}{\partial t}
=
\frac{c}{3r^{2}\rho}
\frac{\partial}{\partial r}
\left(
\frac{r^{2}}{\kappa\rho}
\frac{\partial u}{\partial r}
\right)
+
\epsilon_{\rm heat},
\label{eq:diffusion_equation}
\end{equation}
where $\rho$ is the local mass density, $\kappa$ is the opacity, and $\epsilon_{\rm heat}$ is the specific heating rate. 
The left-hand side describes the time evolution of the radiation energy density and adiabatic cooling, while the first term on the right-hand side represents radiative diffusion.

The shock-heating source is localized near the forward shock and is written as
\begin{equation}
\epsilon_{\rm heat}
=
\frac{L_{\rm shock}}
{4\pi R_{\rm sh}^{2}\rho_{\rm sh,ej}}
\delta(r-R_{\rm sh}),
\label{eq:epsilon_shock}
\end{equation}
where $\rho_{\rm sh,ej}$ is the effective density of the shocked ejecta. 
This term represents the local thermalization of the kinetic energy dissipated at the shock. 
In the thin-shell approximation, we estimate
\begin{equation}
\rho_{\rm sh,ej}
\simeq
\frac{M_{\rm sh}}
{4\pi R_{\rm sh}^{2}\Delta R_{\rm sh}},
\label{eq:rho_shej}
\end{equation}
where $\Delta R_{\rm sh}$ is the shell thickness.

The PWN energy entering the shocked ejecta is treated as an inner boundary flux rather than a volumetric source term. 
At the bubble--shell interface $R_{\rm sh,in}$, the diffusive flux satisfies
\begin{equation}
\frac{L_{\rm diff}}
{4\pi R_{\rm sh,in}^{2}}
=
-
\frac{c}{3\kappa\rho_{\rm sh,ej}}
\left.
\frac{\partial u}{\partial r}
\right|_{r=R_{\rm sh,in}}.
\label{eq:inner_boundary_flux}
\end{equation}
Thus, $L_{\rm diff}$ is coupled to the radiation field at the inner edge of the shocked ejecta and is not prescribed independently.

The outer boundary is set at the photosphere $R_{\rm ph}$, defined by $\tau_{\rm ph}=2/3$. 
Using the Eddington approximation to match the radiation energy density to the diffusive flux, we impose
\begin{equation}
u(R_{\rm ph},t)
=
-
\frac{4}{3\kappa\rho}
\left.
\frac{\partial u}{\partial r}
\right|_{r=R_{\rm ph}} .
\label{eq:photosphere_boundary}
\end{equation}
When the photosphere lies in the unshocked ejecta, we take $\rho=\rho_{\rm ej}(R_{\rm ph},t)$. 
If the photosphere recedes into the shocked shell, or if the shock has reached the outer ejecta boundary, we instead use $\rho=\rho_{\rm sh,ej}$. 
Outside the photosphere, the diffusion approximation breaks down and photons are assumed to free-stream, so that $u(r,t)\propto r^{-2}$.

The solution of Equation~(\ref{eq:diffusion_equation}) gives the spatial and temporal evolution of $u(r,t)$. 
The bolometric luminosity is evaluated at the photosphere using
\begin{equation}
L_{\rm bol}(t)
=
4\pi R_{\rm ph}^{2}(t)\sigma T_{\rm ph}^{4}(t),
\label{eq:stefan_boltzmann}
\end{equation}
with $u(R_{\rm ph},t)=aT_{\rm ph}^{4}(t)$. 
This provides the connection between the computed radiation-energy profile and the observable light curve.

% \begin{figure}[htbp!]
%     \centering
%     \includegraphics[scale = 0.3, trim=0 0 60 65, clip]{figures/L_t.pdf}
% \caption{
% Energy-budget evolution for a representative \texttt{TransFit-MAG} model. 
% The solid curve shows $L_{\rm bol}$, and the other curves show $L_{\rm shock}$, $L_{\rm sd}$, $L_{\rm work}$, $L_{\rm diff}$, and $\mathd E_{\rm pwn}/\mathd t$. 
% The mismatch between $L_{\rm bol}$ and the instantaneous power sources reflects energy storage, adiabatic degradation, and delayed photon diffusion in the expanding ejecta. 
% The vertical dotted line marks the epoch when the external ejecta become optically thin.
% }
% \label{fig:energy_budget}
% \end{figure}

\section{Light-curve Morphologies in Terms of Two Characteristic Timescales}
\label{sec:morphology_timescales}

In \texttt{TransFit-MAG}, bolometric light curves are calculated by coupling the time-dependent radiation-diffusion equation to the dynamical evolution of a magnetar-driven shock. 
Rather prescribing an Arnett-style response to the magnetar spin-down luminosity, 
the model self-consistently tracks the evolving temperature and radiation-energy distributions within the expanding ejecta. As the PWN shock propagates outward, it deposits thermal energy at a moving heating front. Radiation escaping at early times is primarily generated in the shocked layer near this front, producing a potential shock-powered emission component. Conversely, at later times, radiation deposited by the magnetar deeper within the ejecta diffuses through the bulk mass, powering the broad main peak.

This physical picture naturally frames the light-curve morphology around two characteristic timescales: the shock-breakout timescale, $t_{\rm sbo}$, and the main peak timescale, $t_{\rm sn,pk}$. The former marks when photons from the shocked layer can efficiently escape, while the latter dictates when energy deposited in the bulk ejecta emerges as the primary magnetar-powered maximum.

The early shock-powered component peaks when the effective breakout condition is met:
\begin{equation}
\tau_{\rm sbo}
=
\frac{c}{\zeta_{\rm sbo} v_{\rm sh}},
\label{eq:tau_sbo_morphology}
\end{equation}
where $v_{\rm sh}$ is the shock velocity and $\zeta_{\rm sbo} \simeq 2$ is a numerical calibration factor accounting for the delay between shock heating and the observed luminosity maximum. Combining this optical-depth condition with the shock dynamics and the ejecta density profile yields:
\begin{equation}
t_{\rm sbo}
\propto
\left(
\frac{E_{\rm SN}}{E_{\rm m}}
\right)^{2/3}
t_{\rm d}^{1/3}
t_{\rm sd}^{2/3}.
\label{eq:tsbo_general_morphology}
\end{equation}
Thus, $t_{\rm sbo}$ represents a weighted geometric mean of the diffusion and spin-down timescales, modulated by the energy ratio $E_{\rm SN}/E_{\rm m}$. Within this scaling framework, a larger magnetar energy drives a faster PWN shock, shifting breakout to earlier times, whereas a longer spin-down timescale delays it.

Following shock breakout, the deeper magnetar/PWN energy deposition emerges via radiative diffusion. The characteristic timescale for this broad main peak, $t_{\rm sn,pk}$, is governed by the effective diffusion time through the bulk ejecta and scales as:
\begin{equation}
t_{\rm sn,pk}
\propto
\left(
\frac{\kappa M_{\rm ej}}{v_{\rm ej}}
\right)^{1/2}.
\label{eq:tsn_scaling_morphology}
\end{equation}
This indicates that the main peak is primarily controlled by the ejecta mass, opacity, and characteristic velocity. Increases in $M_{\rm ej}$ or $\kappa$ delay the main peak, while a higher $v_{\rm ej}$ enables the stored radiation to escape more rapidly.

To provide a numerical benchmark, we express these timescales in scaled units. The main peak timescale is:
\begin{equation}
t_{\rm sn,pk} \simeq 68.8\, M_{\rm ej,5}^{3/4} E_{\rm SN,51}^{-1/4} \kappa_{0.2}^{1/2} ~{\rm days},
\label{eq:tsn_scaled_morphology}
\end{equation}
where $E_{\rm SN,51} = E_{\rm SN}/10^{51}\,{\rm erg}$, $M_{\rm ej,5} = M_{\rm ej}/5M_\odot$, and $\kappa_{0.2} = \kappa/(0.2\,{\rm cm^2\,g^{-1}})$. Applying the same normalization to the shock-breakout timescale yields:
\begin{equation}
t_{\rm sbo} \simeq 16.9\, M_{\rm ej,5}^{1/4} E_{\rm SN,51}^{7/12} E_{\rm m,51}^{-2/3} t_{\rm sd,5}^{2/3} \kappa_{0.2}^{1/6} ~{\rm days},
\label{eq:tsbo_scaled_morphology}
\end{equation}
where $E_{\rm m,51} = E_{\rm m}/10^{51}\,{\rm erg}$ and $t_{\rm sd,5} = t_{\rm sd}/5\,{\rm days}$. The theoretical temporal separation between the early shock-powered peak and the broad magnetar-powered peak is therefore defined by their ratio:
\begin{equation}
\frac{t_{\rm sn,pk}}{t_{\rm sbo}} \simeq 4.08\, M_{\rm ej,5}^{1/2} E_{\rm m,51}^{2/3} E_{\rm SN,51}^{-5/6} t_{\rm sd,5}^{-2/3} \kappa_{0.2}^{1/3}.
\label{eq:ratio_morphology}
\end{equation}

In practice, resolving a distinct double-peaked light curve observationally requires a more pronounced delay, parameterized as:
\begin{equation}
\frac{t_{\rm sn,pk}}{t_{\rm sbo}}
\gtrsim \chi_t,
\qquad
\chi_t \simeq 3 \text{--} 5 .
\label{eq:double_peak_time_separation}
\end{equation}
If this ratio approaches unity, the shock-breakout emission blends with the rising main light curve, manifesting only as an early shoulder or kink. Substituting the scaled timescales into this criterion yields the necessary condition for temporal separation:
\begin{equation}
E_{\mathrm{m},51}
\gtrsim
0.63
\left(\frac{\chi_t}{3}\right)^{3/2}
M_{\mathrm{ej},5}^{-3/4}
E_{\mathrm{SN},51}^{5/4}
t_{\mathrm{sd},5}
\kappa_{0.2}^{-1/2}.
\label{eq:double_peak_energy_condition_time}
\end{equation}
This condition directly demonstrates that higher magnetar energy accelerates the PWN shock (triggering earlier breakout), while greater ejecta mass or opacity delays the main peak. Both effects inflate the $t_{\rm sn,pk}/t_{\rm sbo}$ ratio, thereby favoring a well-separated double-peaked morphology.

However, temporal separation alone is insufficient; the shock-powered component must also outshine the underlying diffusive luminosity at the time of breakout:
\begin{equation}
L_{\mathrm{sbo}}
\gtrsim
L_{\mathrm{sn}}(t_{\mathrm{sbo}}) .
\label{eq:double_peak_luminosity_condition}
\end{equation}
The breakout peak need not eclipse the maximum luminosity of the main peak, $L_{\rm sn,pk}$, but it must dominate the rising magnetar-powered component at $t_{\mathrm{sbo}}$. Assuming the early evolution of the main component follows a power-law rise,
\begin{equation}
L_{\mathrm{sn}}(t_{\mathrm{sbo}})
\simeq
f_{\mathrm{th, early}}
L_{\rm sn,pk}
\left(
\frac{t_{\mathrm{sbo}}}{t_{\rm sn,pk}}
\right)^q,
\qquad
q \simeq 1.5 \text{--} 2 ,
\label{eq:early_diffusive_luminosity}
\end{equation}
the required luminosity contrast is:
\begin{equation}
\frac{L_{\mathrm{sbo}}}{L_{\rm sn,pk}}
\gtrsim
f_{\mathrm{th, early}}
\left(
\frac{t_{\mathrm{sbo}}}{t_{\rm sn,pk}}
\right)^q ,
\label{eq:double_peak_contrast_condition}
\end{equation}
where $f_{\mathrm{th, early}} \leq 1$ represents the effective early-time thermalization efficiency of the magnetar/PWN input. This condition dictates that even a relatively weak breakout component can manifest as a distinct first peak if it precedes the main maximum significantly. For instance, with $t_{\rm sn,pk}/t_{\mathrm{sbo}} \sim 4$ and $q \simeq 2$, a breakout component reaching just a few percent of the main-peak luminosity remains discernible, provided the early diffusive emission is suitably suppressed.

\begin{figure}[htbp!]
\centering
\includegraphics[width=0.95\columnwidth, trim=0 0 60 65, clip]{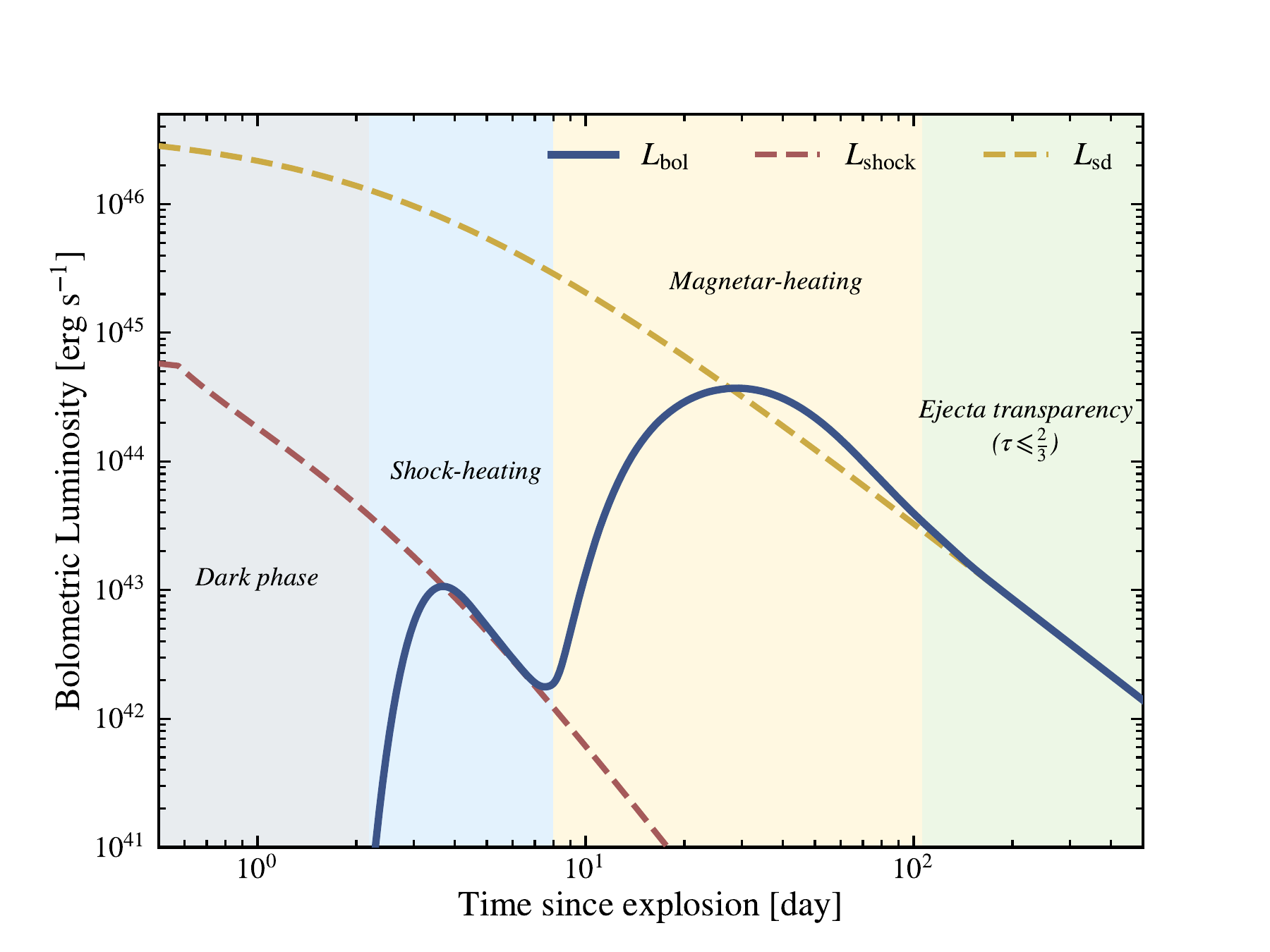}\\[-0.2em]
\includegraphics[width=0.95\columnwidth, trim=0 0 60 65, clip]{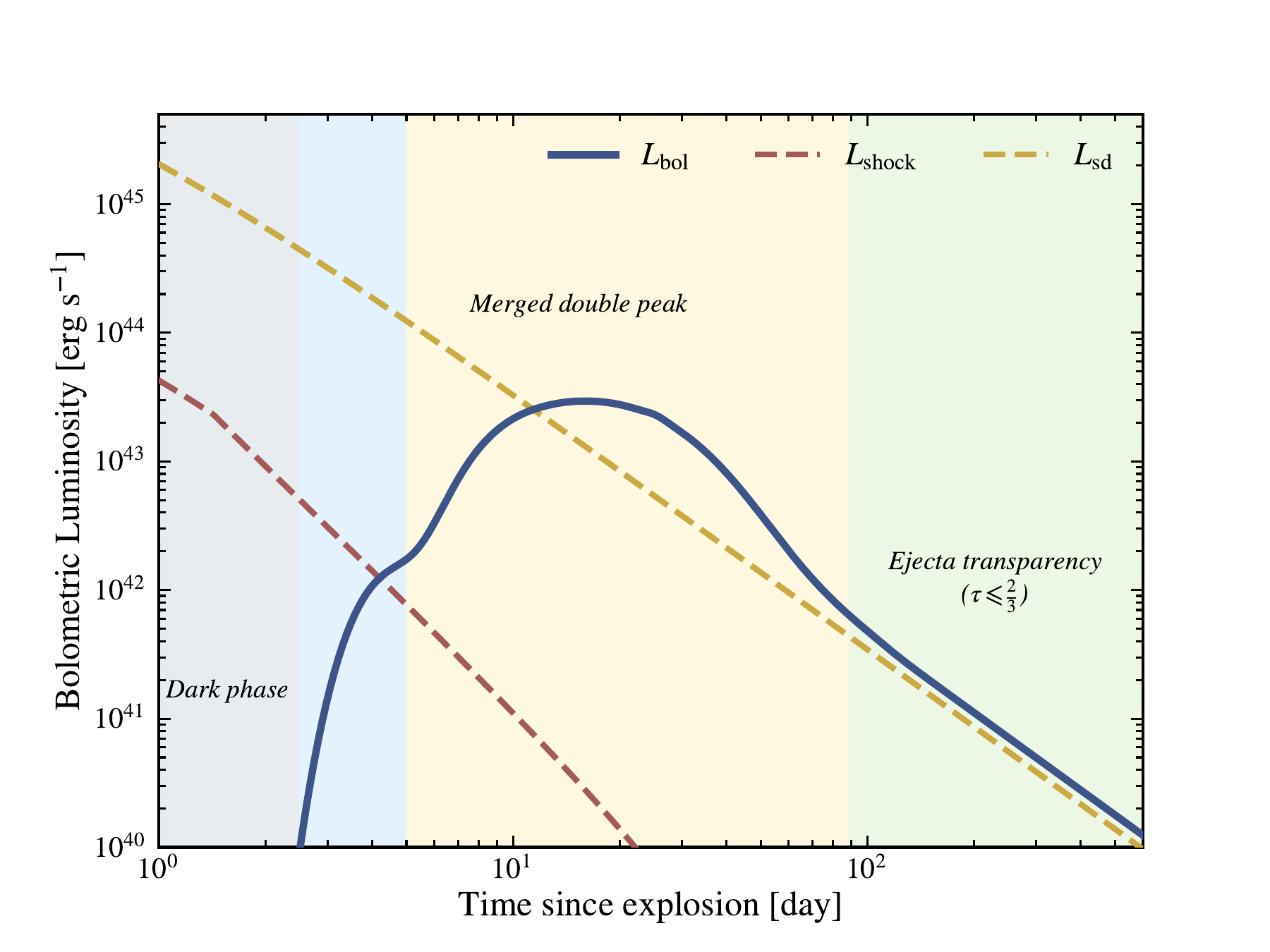}\\[-0.2em]
\includegraphics[width=0.95\columnwidth, trim=0 0 60 65, clip]{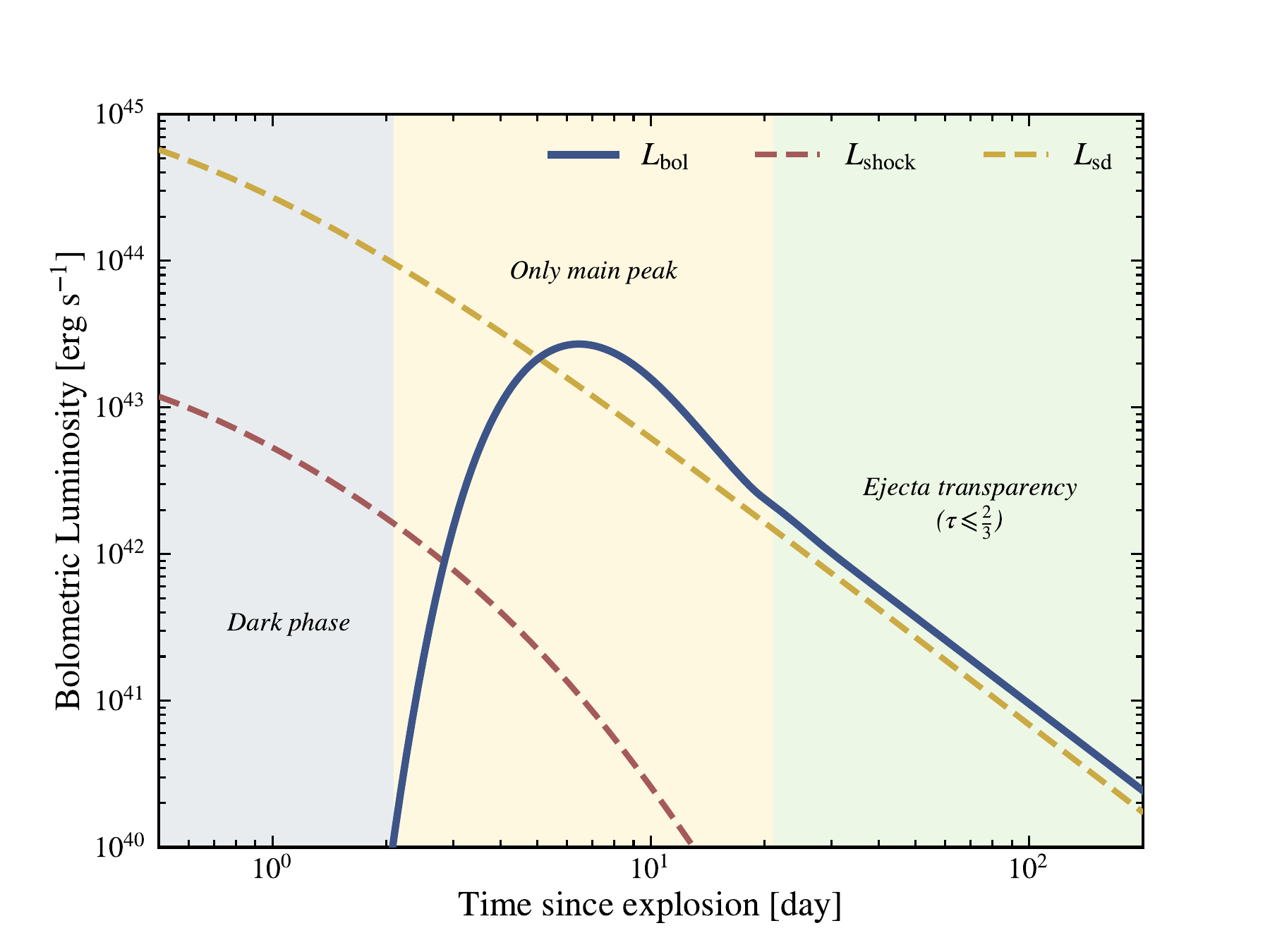}
\caption{ Representative bolometric light curves from \texttt{TransFit-MAG}, illustrating three characteristic light-curve morphologies. From top to bottom, the panels show a well-separated double peak, a partially merged double peak, and a single broad peak. The blue solid curve shows the emergent bolometric luminosity $L_{\rm bol}$, while the brown and gold dashed curves show the instantaneous shock-heating power $L_{\rm shock}$ and the magnetar spin-down luminosity $L_{\rm sd}$, respectively. The shaded regions mark the main evolutionary stages: the initial dark phase, the shock-heating dominated phase, the magnetar/PWN-heating dominated diffusion phase, and the late optically thin phase when the bulk ejecta becomes transparent. The transition from a clear double peak to a single broad peak is controlled mainly by the temporal separation between $t_{\rm sbo}$ and $t_{\rm SN,pk}$, together with the luminosity contrast between $L_{\rm sbo}$ and the underlying diffusive luminosity $L_{\rm SN}(t_{\rm sbo})$. }
\label{fig:lc_morphology}
\end{figure}

Figure~\ref{fig:lc_morphology} illustrates how the above criteria map onto the different light-curve morphologies obtained with \texttt{TransFit-MAG}. In the well-separated double-peaked case, the shock breakout occurs sufficiently earlier than the main diffusion maximum, $t_{\rm sbo}\ll t_{\rm SN,pk}$, and the shock-powered luminosity exceeds the rising diffusive component at the breakout time, $L_{\rm sbo}\gtrsim L_{\rm SN}(t_{\rm sbo})$. The first maximum is therefore produced by radiation released from the shocked outer layers, whereas the second, broader maximum is powered by magnetar/PWN energy that diffuses out from deeper ejecta layers. The luminosity minimum between the two peaks appears after the shock-heating power has declined but before the main magnetar-powered component reaches its maximum.

The middle panel shows an intermediate case in which the two characteristic timescales are less clearly separated. Although the shock-heating component still modifies the early light curve, the magnetar-powered diffusion component rises before a deep luminosity valley can form. The result is a partially merged double peak, or an early shoulder on the rising part of the main peak. This morphology corresponds to cases in which $t_{\rm sbo}$ is only moderately smaller than $t_{\rm SN,pk}$, or in which $L_{\rm sbo}$ is comparable to, but not much larger than, $L_{\rm SN}(t_{\rm sbo})$.

The bottom panel represents the single-peaked limit. In this case, the shock-powered emission is either too weak or too close in time to the broad diffusion peak to appear as a separate component. Equivalently, the condition $L_{\rm sbo}\gtrsim L_{\rm SN}(t_{\rm sbo})$ is not satisfied, or the ratio $t_{\rm SN,pk}/t_{\rm sbo}$ is too small. The observed luminosity is then dominated by the main magnetar-powered diffusion component, and the breakout signature is hidden in the early rise. This sequence demonstrates that the appearance of a double peak is not determined solely by the absolute value of the shock-heating power. Instead, it is controlled by the competition between the shock-breakout emission and the underlying diffusive luminosity at $t_{\rm sbo}$. A clear double peak requires both a sufficiently early and dynamically strong shock breakout and a delayed rise of the main magnetar-powered component. Conversely, efficient early thermalization of the magnetar power raises $L_{\rm SN}(t_{\rm sbo})$, making the breakout component easier to hide and driving the morphology toward a merged or single-peaked light curve.

% \begin{deluxetable}{lccc}
% \tabletypesize{\footnotesize} \tablewidth{0.95\columnwidth}
% \tablecaption{
% Model parameters, priors, and best-fit values for the LSQ14bdq light-curve modeling.
% \label{tab:fit_priors}
% }
% \tablehead{
% \colhead{Parameter} &
% \colhead{Prior} &
% \colhead{Range} &
% \colhead{Best Fit}
% }
% \startdata
% $E_{\rm SN}\ (10^{51}\,{\rm erg})$        & Log-uniform & $[0.1,\,10.0]$   & $6.98$ \\
% $M_{\rm ej}\ (M_\odot)$                  & Uniform     & $[0.1,\,40.0]$   & $37.54$ \\
% $E_{\rm m}\ (10^{51}\,{\rm erg})$         & Log-uniform & $[0.1,\,25.1]$   & $25.1$ \\
% $t_{\rm sd}\ ({\rm days})$               & Uniform     & $[0.1,\,50.0]$   & $30.10$ \\
% $\kappa\ ({\rm cm^{2}\,{\rm g}^{-1}})$   & Uniform     & $[0.1,\,0.34]$   & $0.340$ \\
% $t_{\rm shift}\ ({\rm days})$            & Uniform     & $[0.0,\,30.0]$   & $27.10$ \\
% \enddata
% \tablecomments{
% The energy parameters are sampled with log-uniform priors over the listed ranges, while the remaining parameters use uniform priors. 
% The best-fit values of $E_{\rm m}$ and $\kappa$ reach the upper boundaries of their adopted prior ranges.
% }
% \end{deluxetable}

\section{Application to LSQ14bdq}
\label{sec:lsq14bdq_fit}

\begin{figure*}[t]
\centering
\includegraphics[width=0.48\textwidth]{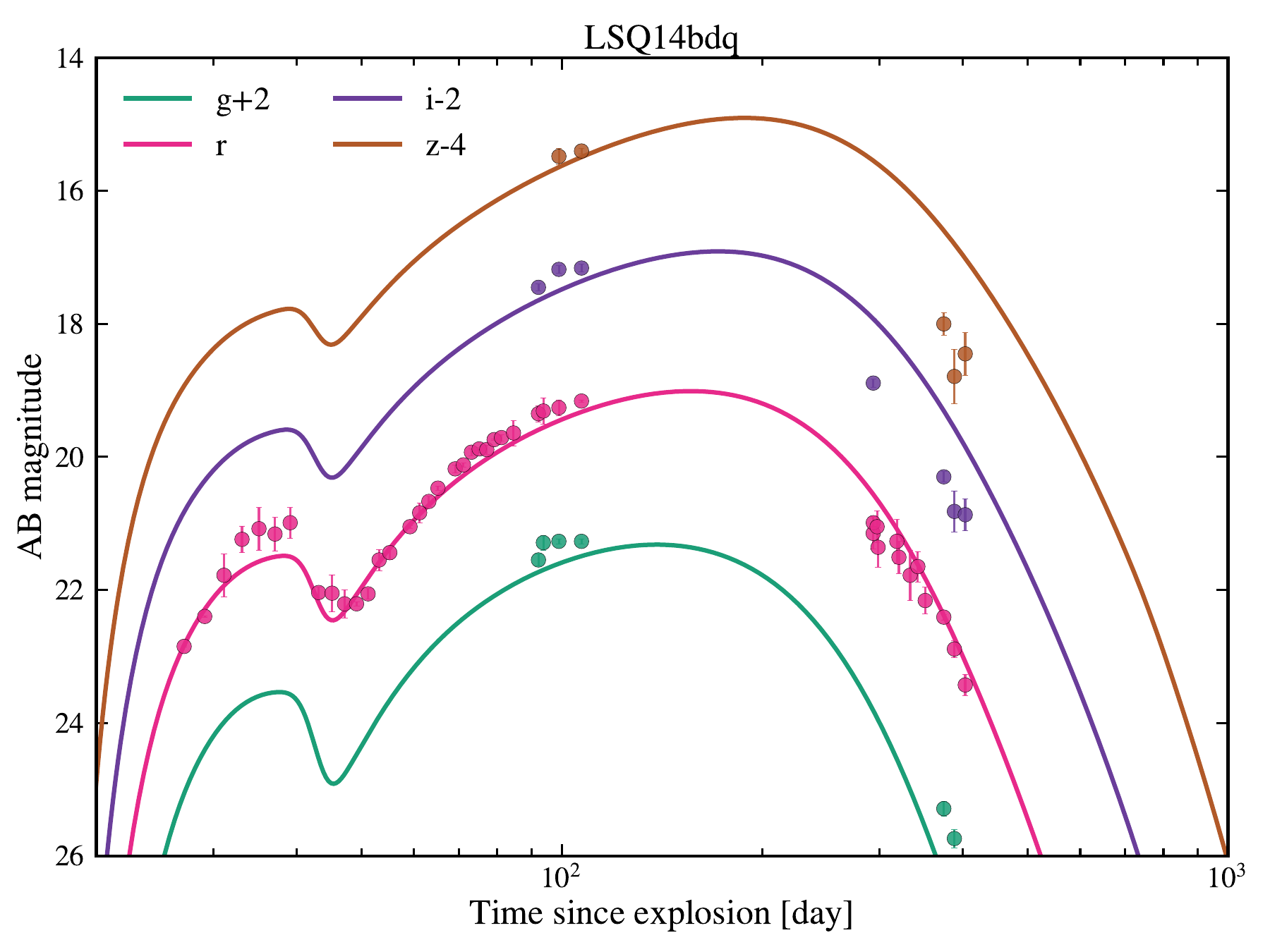}
\hfill
\includegraphics[width=0.48\textwidth]{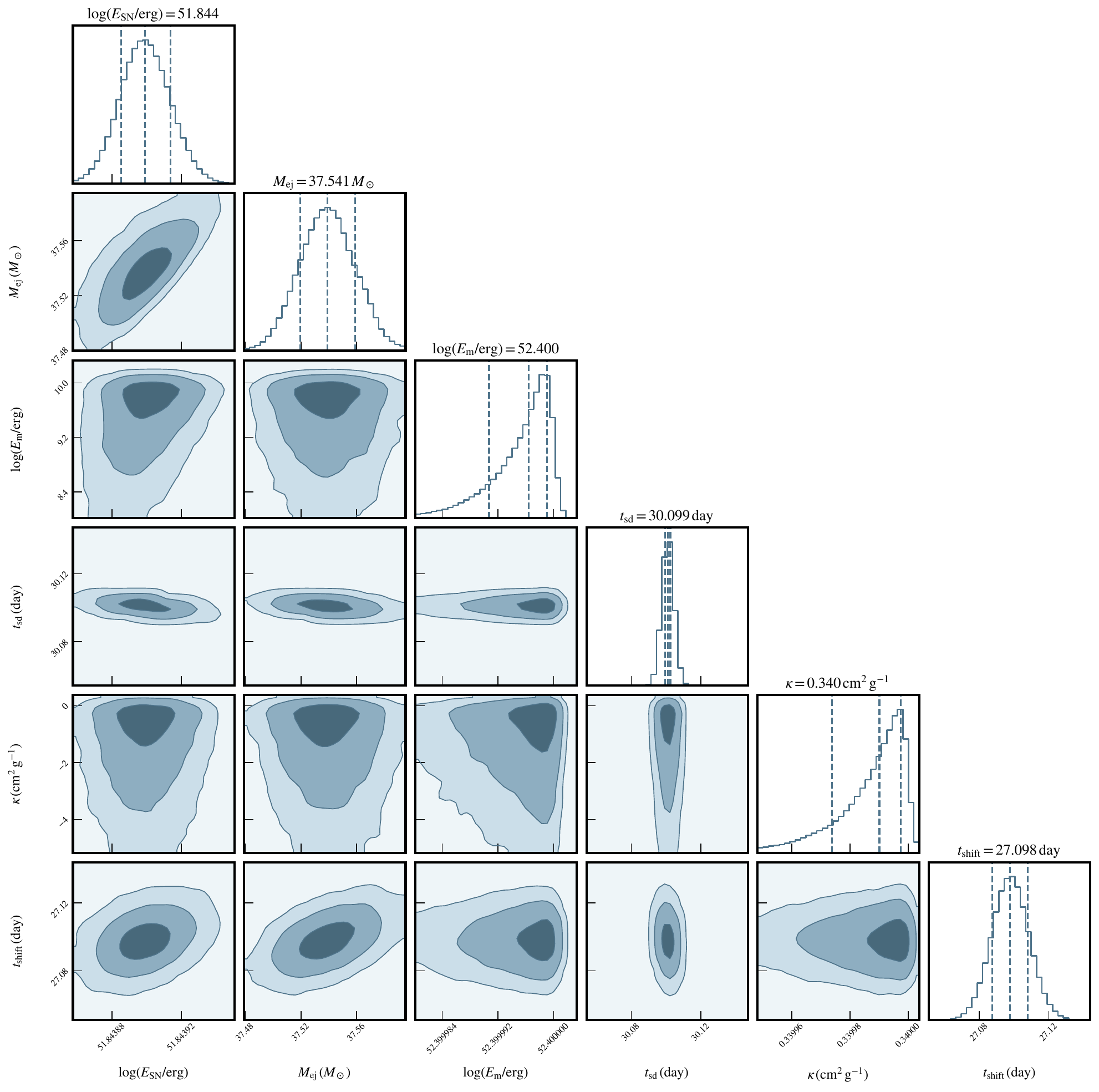}
\caption{
Application of \texttt{TransFit-MAG} to the multiband optical light curves of
LSQ14bdq. The left panel shows the observed $g$, $r$, $i$, and $z$-band
light curves together with the best-fit model. Vertical offsets are applied
for clarity. The right panel shows the posterior distributions of the model
parameters. The red lines indicate the best-fit values, and the dashed lines
mark the marginalized credible intervals. The model reproduces both the early
bump and the broad main peak without imposing an explicit delay time between
magnetar injection and radiative escape.
}
\label{fig:lsq14bdq_fit}
\end{figure*}

LSQ14bdq provides a useful test case for \texttt{TransFit-MAG}, because it
exhibits one of the clearest double-peaked optical light curves among
SLSNe-I \citep{Nicholl2015}. 
%Its luminous early bump and broader main peak offer a direct opportunity to test whether the coupled PWN--shock--diffusion framework can reproduce both components within a single engine-powered scenario. 
In particular, LSQ14bdq is well suited for testing whether the
temporal separation between the early bump and the main maximum can arise
from the evolving optical depth and radiation-energy distribution, rather
than from an externally imposed delay between magnetar injection and photon
escape. Then, we apply \texttt{TransFit-MAG} to the multiband optical light curves of
LSQ14bdq using the data compiled by \citet{Nicholl2015}. The resulting fit
and posterior distributions are shown in Figure~\ref{fig:lsq14bdq_fit}. In our interpretation, the early bump is produced mainly by the escape of
shock-heated radiation from the outer ejecta, whereas the main peak is powered
by delayed diffusion of magnetar/PWN energy deposited deeper inside the
ejecta. The luminosity minimum between the two components appears naturally
after the shock-heating power has declined but before the main
magnetar-powered diffusion component reaches its maximum.

%The best-fit values and a comparison with parameters inferred in previous studies is given in Table~\ref{tab:lsq14bdq_parameter_comparison}. 
The best-fit parameters are
$E_{\rm SN}=6.98\times10^{51}\,{\rm erg}$,
$M_{\rm ej}=37.54\,M_\odot$,
$E_{\rm m}=2.51\times10^{52}\,{\rm erg}$,
$t_{\rm sd}=30.10\,{\rm days}$,
$\kappa=0.34\,{\rm cm^2\,g^{-1}}$, and
$t_{\rm shift}=27.10\,{\rm days}$. A key feature of our fit is that no explicit delay time,
$t_{\rm delay}$, is introduced between magnetar spin-down and radiative
escape. Previous semi-analytic treatments often invoke such a delay, or an
early suppression of magnetar thermalization, to separate the early bump from
the main peak \citep[e.g.,][]{kasen2016,liu2021}. In
\texttt{TransFit-MAG}, the separation is instead generated by the evolving
optical depth, shock-heating location, and internal radiation-energy
distribution. 
%The effective delay between energy injection and photon escape is therefore an output of the coupled dynamical--diffusion calculation rather than an independent fitting parameter.
%Figure~\ref{fig:lsq14bdq_fit} shows that the model captures the two main features of LSQ14bdq 
The large ejecta mass and relatively high
opacity increase the effective diffusion time and help broaden the main peak.
The large magnetar rotational-energy reservoir provides the energy required
for the high luminosity, while the relatively long spin-down timescale keeps
the central engine active through the rise of the main peak. These parameters
therefore place LSQ14bdq in the regime where an early shock-powered component
can be separated from a delayed magnetar-powered diffusion maximum.

In applying the model to LSQ14bdq, we find that the assumed ejecta density
structure has a significant effect on the early peak. For a standard broken
power-law density profile, the shock-heated layer remains buried beneath an
optically thick outer envelope during the early phase. A substantial fraction
of the deposited thermal energy is then degraded by adiabatic expansion and
smoothed by diffusion before reaching the photosphere, producing a relatively
faint first peak. By contrast, an effective single power-law density profile
places the early photosphere closer to the outer ejecta boundary and reduces
the diffusion time of the shock-heated radiation. The early emission is
therefore more concentrated in time, making it easier to reproduce the narrow
and luminous first peak of LSQ14bdq. We emphasize that this treatment should
be regarded as an effective description of the density structure relevant for
the early photon escape, rather than as a unique reconstruction of the ejecta
profile.

\section{Conclusion and Discussions}
\label{sec:discussion_conclusion}

We have developed \texttt{TransFit-MAG}, a time-dependent radiative-diffusion framework for engine-powered transients with magnetar-driven shocks. 
The model couples the \texttt{TransFit} diffusion solver to the dynamics of a magnetar-inflated PWN and its forward shock in homologously expanding ejecta. 
The spin-down power inflates a hot PWN bubble, which performs mechanical work on the swept-up shell and drives shock heating. 
The deposited radiation is then stored, degraded by adiabatic expansion, and released through time-dependent diffusion. 
%This treatment follows the internal-energy distribution, photospheric evolution, and emergent luminosity without imposing a fixed diffusion time or an artificial delay between energy injection and photon escape.

The shock evolution provides a useful check on the dynamical treatment. 
While the shock remains in the shallow inner ejecta and radiative leakage is negligible, the numerical solution approaches the analytic self-similar behavior. 
After the shock reaches the density break and enters the steep outer ejecta, the evolution departs from this limit. 
At later times, radiative diffusion removes energy from the PWN--shocked-ejecta system, reducing the bubble pressure and weakening the shell acceleration. 
The emergent luminosity therefore does not directly follow either the instantaneous spin-down luminosity $L_{\rm sd}$ or the shock-heating power $L_{\rm shock}$, but is delayed and smoothed by diffusion through the expanding ejecta.

Our calculations show that the competition between shock-heated radiation and delayed magnetar/PWN diffusion naturally produces a range of light-curve morphologies. 
If the shock-heated radiation escapes early enough and remains sufficiently luminous, the light curve shows a distinct first peak. 
If this component overlaps with the rising magnetar-powered emission, it appears as a shoulder or partially merged bump. 
If the shock is weak or deeply embedded, diffusion and adiabatic losses suppress the early component, leaving a single broad peak. 
Thus, the presence of an early bump is not set by the magnetar engine alone, but by the coupled engine--shock--diffusion evolution and the optical-depth structure of the ejecta.

This work is complementary to population-level studies of magnetar-powered transients. 
Such studies suggest that magnetar engines may operate in several classes of stripped-envelope explosions, including SLSNe, FBOTs, and SNe Ic-BL \citep{yu2017,liu2022,zhu2026}. 
While population modeling constrains distributions of ejecta masses, spin periods, magnetic fields, and explosion energies, \texttt{TransFit-MAG} addresses the time-dependent energy transport in individual events. 
It therefore provides a way to connect spatially separated shock-heated and engine-heated components to the observed early peak, shoulder-like bump, or broad main peak.

In summary, \texttt{TransFit-MAG} provides a self-consistent framework for connecting central-engine injection, PWN-driven shock heating, and time-dependent radiative diffusion in expanding supernova ejecta. 
The model shows that early bumps in engine-powered transients can arise naturally from coupled shock--diffusion evolution, without requiring an imposed delay between engine injection and photon escape. 
Together with population studies, this framework provides a useful tool for interpreting early observations from high-cadence surveys and for constraining the physical conditions of engine-powered stripped-envelope explosions. 

%As an illustrative application, we fitted the multiband optical light curves of the double-peaked SLSN-I LSQ14bdq. The model reproduces both the early bump and the broader main peak without introducing an explicit delay time. In our interpretation, the early bump is associated with the diffusion of shock-heated radiation, whereas the main peak is powered by delayed diffusion of magnetar/PWN energy. The inferred large ejecta mass should be interpreted with caution, because it is degenerate with the adopted opacity, density profile, and photospheric emission prescription.

Several limitations remain. 
First, the model adopts a one-dimensional thin-shell treatment of the PWN-driven shock and does not capture multidimensional instabilities, shell fragmentation, or mixing at the bubble--ejecta interface. 
Such effects can lead to wind-bubble blowout and may modify the early UV/optical emission \citep{Chen2026}. 
Second, we use a gray-opacity diffusion approximation, whereas wavelength-dependent radiative transfer is required for detailed color and spectral modeling. 
Third, after the ejecta become optically thin, non-thermal PWN emission may contribute significantly to the late-time radiation. 
Hard X-ray and gamma-ray emission from embedded PWNe may provide an independent diagnostic of fast-spinning newborn pulsars in stripped-envelope supernovae \citep{Kashiyama2016}. 
Future multiwavelength observations and multidimensional radiation-hydrodynamic simulations will be important for testing the thin-shell approximation, the energy-deposition treatment, and the inferred ejecta properties.

\begin{acknowledgements}
This work was supported by the National Natural Science Foundation of China
(grant Nos. 12303047 and 12393811), the Natural Science Foundation of Hubei Province
(grant No. 2023AFB321), and the National Key R\&D Program of China
(grant No. 2021YFA0718500).
\end{acknowledgements}

\bibliography{sample701}{}

@ARTICLE{Chen2026,
       author = {{Chen}, Mingxi and {Kashiyama}, Kazumi and {Sato}, Masato},
        title = "{Blowouts of Nascent Wind Bubbles in Pulsar-driven Supernovae}",
      journal = {\apj},
         year = 2026,
        month = apr,
       volume = {1001},
       number = {1},
          eid = {59},
        pages = {59},
          doi = {10.3847/1538-4357/ae4c58},
archivePrefix = {arXiv},
       eprint = {2601.09552},
 primaryClass = {astro-ph.HE},
       adsurl = {https://ui.adsabs.harvard.edu/abs/2026ApJ..1001...59C}
}

@ARTICLE{Lin2021,
       author = {{Lin}, Weili and {Wang}, Xiaofeng and {Wang}, Lingjun and {Dai}, Zigao},
        title = "{Supernova Luminosity Powered by Magnetar-Disk System}",
      journal = {\apjl},
         year = 2021,
        month = jun,
       volume = {914},
       number = {1},
          eid = {L2},
        pages = {L2},
          doi = {10.3847/2041-8213/ac004a},
archivePrefix = {arXiv},
       eprint = {2105.05512},
 primaryClass = {astro-ph.HE},
       adsurl = {https://ui.adsabs.harvard.edu/abs/2021ApJ...914L...2L}
}

@ARTICLE{Piro2011,
       author = {{Piro}, Anthony L. and {Ott}, Christian D.},
        title = "{Supernova Fallback onto Magnetars and Propeller-powered Supernovae}",
      journal = {\apj},
         year = 2011,
        month = aug,
       volume = {736},
       number = {2},
          eid = {108},
        pages = {108},
          doi = {10.1088/0004-637X/736/2/108},
archivePrefix = {arXiv},
       eprint = {1104.0252},
 primaryClass = {astro-ph.HE},
       adsurl = {https://ui.adsabs.harvard.edu/abs/2011ApJ...736..108P}
}

@ARTICLE{Moriya2019,
       author = {{Moriya}, Takashi J. and {M{\"u}ller}, Bernhard and {Chan}, Conrad and {Heger}, Alexander and {Blinnikov}, Sergei I.},
        title = "{Fallback Accretion-powered Supernova Light Curves Based on a Neutrino-driven Explosion Simulation of a 40 M $_{☉}$ Star}",
      journal = {\apj},
         year = 2019,
        month = jul,
       volume = {880},
       number = {1},
          eid = {21},
        pages = {21},
          doi = {10.3847/1538-4357/ab2643},
archivePrefix = {arXiv},
       eprint = {1906.00153},
 primaryClass = {astro-ph.HE},
       adsurl = {https://ui.adsabs.harvard.edu/abs/2019ApJ...880...21M}
}

@ARTICLE{Dexter2013,
       author = {{Dexter}, Jason and {Kasen}, Daniel},
        title = "{Supernova Light Curves Powered by Fallback Accretion}",
      journal = {\apj},
         year = 2013,
        month = jul,
       volume = {772},
       number = {1},
          eid = {30},
        pages = {30},
          doi = {10.1088/0004-637X/772/1/30},
archivePrefix = {arXiv},
       eprint = {1210.7240},
 primaryClass = {astro-ph.HE},
       adsurl = {https://ui.adsabs.harvard.edu/abs/2013ApJ...772...30D}
}

@ARTICLE{Inserra2019,
       author = {{Inserra}, C.},
        title = "{Observational properties of extreme supernovae}",
      journal = {Nature Astronomy},
         year = 2019,
        month = aug,
       volume = {3},
        pages = {697-705},
          doi = {10.1038/s41550-019-0854-4},
archivePrefix = {arXiv},
       eprint = {1908.02314},
 primaryClass = {astro-ph.HE},
       adsurl = {https://ui.adsabs.harvard.edu/abs/2019NatAs...3..697I}
}

@ARTICLE{Nicholl2015,
       author = {{Nicholl}, M. and {Smartt}, S.~J. and {Jerkstrand}, A. and {Sim}, S.~A. and {Inserra}, C. and {Anderson}, J.~P. and {Baltay}, C. and {Benetti}, S. and {Chambers}, K. and {Chen}, T.-W. and {Elias-Rosa}, N. and {Feindt}, U. and {Flewelling}, H.~A. and {Fraser}, M. and {Gal-Yam}, A. and {Galbany}, L. and {Huber}, M.~E. and {Kangas}, T. and {Kankare}, E. and {Kotak}, R. and {Kr{\"u}hler}, T. and {Maguire}, K. and {McKinnon}, R. and {Rabinowitz}, D. and {Rostami}, S. and {Schulze}, S. and {Smith}, K.~W. and {Sullivan}, M. and {Tonry}, J.~L. and {Valenti}, S. and {Young}, D.~R.},
        title = "{LSQ14bdq: A Type Ic Super-luminous Supernova with a Double-peaked Light Curve}",
      journal = {\apjl},
         year = 2015,
        month = jul,
       volume = {807},
       number = {1},
          eid = {L18},
        pages = {L18},
          doi = {10.1088/2041-8205/807/1/L18},
archivePrefix = {arXiv},
       eprint = {1505.01078},
 primaryClass = {astro-ph.SR},
       adsurl = {https://ui.adsabs.harvard.edu/abs/2015ApJ...807L..18N}
}

@ARTICLE{Nicholl2016,
       author = {{Nicholl}, M. and {Smartt}, S.~J.},
        title = "{Seeing double: the frequency and detectability of double-peaked superluminous supernova light curves}",
      journal = {\mnras},
         year = 2016,
        month = mar,
       volume = {457},
       number = {1},
        pages = {L79-L83},
          doi = {10.1093/mnrasl/slv210},
archivePrefix = {arXiv},
       eprint = {1511.03740},
 primaryClass = {astro-ph.SR},
       adsurl = {https://ui.adsabs.harvard.edu/abs/2016MNRAS.457L..79N}
}

@ARTICLE{Smith2016,
       author = {{Smith}, M. and {Sullivan}, M. and {D'Andrea}, C.~B. and {Castander}, F.~J. and {Casas}, R. and {Prajs}, S. and {Papadopoulos}, A. and {Nichol}, R.~C. and {Karpenka}, N.~V. and {Bernard}, S.~R. and {Brown}, P. and {Cartier}, R. and {Cooke}, J. and {Curtin}, C. and {Davis}, T.~M. and {Finley}, D.~A. and {Foley}, R.~J. and {Gal-Yam}, A. and {Goldstein}, D.~A. and {Gonz{\'a}lez-Gait{\'a}n}, S. and {Gupta}, R.~R. and {Howell}, D.~A. and {Inserra}, C. and {Kessler}, R. and {Lidman}, C. and {Marriner}, J. and {Nugent}, P. and {Pritchard}, T.~A. and {Sako}, M. and {Smartt}, S. and {Smith}, R.~C. and {Spinka}, H. and {Thomas}, R.~C. and {Wolf}, R.~C. and {Zenteno}, A. and {Abbott}, T.~M.~C. and {Benoit-L{\'e}vy}, A. and {Bertin}, E. and {Brooks}, D. and {Buckley-Geer}, E. and {Carnero Rosell}, A. and {Carrasco Kind}, M. and {Carretero}, J. and {Crocce}, M. and {Cunha}, C.~E. and {da Costa}, L.~N. and {Desai}, S. and {Diehl}, H.~T. and {Doel}, P. and {Estrada}, J. and {Evrard}, A.~E. and {Flaugher}, B. and {Fosalba}, P. and {Frieman}, J. and {Gerdes}, D.~W. and {Gruen}, D. and {Gruendl}, R.~A. and {James}, D.~J. and {Kuehn}, K. and {Kuropatkin}, N. and {Lahav}, O. and {Li}, T.~S. and {Marshall}, J.~L. and {Martini}, P. and {Miller}, C.~J. and {Miquel}, R. and {Nord}, B. and {Ogando}, R. and {Plazas}, A.~A. and {Reil}, K. and {Romer}, A.~K. and {Roodman}, A. and {Rykoff}, E.~S. and {Sanchez}, E. and {Scarpine}, V. and {Schubnell}, M. and {Sevilla-Noarbe}, I. and {Soares-Santos}, M. and {Sobreira}, F. and {Suchyta}, E. and {Swanson}, M.~E.~C. and {Tarle}, G. and {Walker}, A.~R. and {Wester}, W. and {DES Collaboration}},
        title = "{DES14X3taz: A Type I Superluminous Supernova Showing a Luminous, Rapidly Cooling Initial Pre-peak Bump}",
      journal = {\apjl},
         year = 2016,
        month = feb,
       volume = {818},
       number = {1},
          eid = {L8},
        pages = {L8},
          doi = {10.3847/2041-8205/818/1/L8},
archivePrefix = {arXiv},
       eprint = {1512.06043},
 primaryClass = {astro-ph.SR},
       adsurl = {https://ui.adsabs.harvard.edu/abs/2016ApJ...818L...8S}
}

@ARTICLE{Leloudas2012,
       author = {{Leloudas}, G. and {Chatzopoulos}, E. and {Dilday}, B. and {Gorosabel}, J. and {Vinko}, J. and {Gallazzi}, A. and {Wheeler}, J.~C. and {Bassett}, B. and {Fischer}, J.~A. and {Frieman}, J.~A. and {Fynbo}, J.~P.~U. and {Goobar}, A. and {Jel{\'\i}nek}, M. and {Malesani}, D. and {Nichol}, R.~C. and {Nordin}, J. and {{\"O}stman}, L. and {Sako}, M. and {Schneider}, D.~P. and {Smith}, M. and {Sollerman}, J. and {Stritzinger}, M.~D. and {Th{\"o}ne}, C.~C. and {de Ugarte Postigo}, A.},
        title = "{SN 2006oz: rise of a super-luminous supernova observed by the SDSS-II SN Survey}",
      journal = {\aap},
         year = 2012,
        month = may,
       volume = {541},
          eid = {A129},
        pages = {A129},
          doi = {10.1051/0004-6361/201118498},
archivePrefix = {arXiv},
       eprint = {1201.5393},
 primaryClass = {astro-ph.SR},
       adsurl = {https://ui.adsabs.harvard.edu/abs/2012A&A...541A.129L}
}

@ARTICLE{Angus2019,
       author = {{Angus}, C.~R. and {Smith}, M. and {Sullivan}, M. and {Inserra}, C. and {Wiseman}, P. and {D'Andrea}, C.~B. and {Thomas}, B.~P. and {Nichol}, R.~C. and {Galbany}, L. and {Childress}, M. and {Asorey}, J. and {Brown}, P.~J. and {Casas}, R. and {Castander}, F.~J. and {Curtin}, C. and {Frohmaier}, C. and {Glazebrook}, K. and {Gruen}, D. and {Gutierrez}, C. and {Kessler}, R. and {Kim}, A.~G. and {Lidman}, C. and {Macaulay}, E. and {Nugent}, P. and {Pursiainen}, M. and {Sako}, M. and {Soares-Santos}, M. and {Thomas}, R.~C. and {Abbott}, T.~M.~C. and {Avila}, S. and {Bertin}, E. and {Brooks}, D. and {Buckley-Geer}, E. and {Burke}, D.~L. and {Carnero Rosell}, A. and {Carretero}, J. and {da Costa}, L.~N. and {De Vicente}, J. and {Desai}, S. and {Diehl}, H.~T. and {Doel}, P. and {Eifler}, T.~F. and {Flaugher}, B. and {Fosalba}, P. and {Frieman}, J. and {Garc{\'\i}a-Bellido}, J. and {Gruendl}, R.~A. and {Gschwend}, J. and {Hartley}, W.~G. and {Hollowood}, D.~L. and {Honscheid}, K. and {Hoyle}, B. and {James}, D.~J. and {Kuehn}, K. and {Kuropatkin}, N. and {Lahav}, O. and {Lima}, M. and {Maia}, M.~A.~G. and {March}, M. and {Marshall}, J.~L. and {Menanteau}, F. and {Miller}, C.~J. and {Miquel}, R. and {Ogando}, R.~L.~C. and {Plazas}, A.~A. and {Romer}, A.~K. and {Sanchez}, E. and {Schindler}, R. and {Schubnell}, M. and {Sobreira}, F. and {Suchyta}, E. and {Swanson}, M.~E.~C. and {Tarle}, G. and {Thomas}, D. and {Tucker}, D.~L. and {DES Collaboration}},
        title = "{Superluminous supernovae from the Dark Energy Survey}",
      journal = {\mnras},
         year = 2019,
        month = aug,
       volume = {487},
       number = {2},
        pages = {2215-2241},
          doi = {10.1093/mnras/stz1321},
archivePrefix = {arXiv},
       eprint = {1812.04071},
 primaryClass = {astro-ph.HE},
       adsurl = {https://ui.adsabs.harvard.edu/abs/2019MNRAS.487.2215A}
}

@article{kasen2016,
  author = {{Kasen}, D. and {Metzger}, B. D. and {Bildsten}, L.},
  title = "{Magnetar-driven Shock Breakout and Double-peaked Supernova Light Curves}",
  journal = {\apj},
  year = {2016},
  volume = {821},
  pages = {36},
  doi = {10.3847/0004-637x/821/1/36}
}

@article{suzuki2021,
  author = {{Suzuki}, A. and {Maeda}, K.},
  title = "{Two-dimensional Radiation-hydrodynamic Simulations of Supernova Ejecta with a Central Power Source}",
  journal = {\apj},
  year = {2021},
  volume = {908},
  pages = {217},
  doi = {10.3847/1538-4357/abd54c}
}

@article{dai1998,
  author = {{Dai}, Z. G. and {Lu}, T.},
  title = "{$\gamma$-Ray Bursts and Afterglows from Rotating Strange Stars and Neutron Stars}",
  journal = {\prl},
  year = {1998},
  volume = {81},
  pages = {4301--4304},
  doi = {10.1103/physrevlett.81.4301}
}

@article{kasen2010,
  author = {{Kasen}, D. and {Bildsten}, L.},
  title = "{Supernova Light Curves Powered by Young Magnetars}",
  journal = {\apj},
  year = {2010},
  volume = {717},
  pages = {245--249},
  doi = {10.1088/0004-637x/717/1/245}
}

@article{chevalier1992,
  author = {{Chevalier}, R. A. and {Fransson}, C.},
  title = "{Pulsar Nebulae in Supernovae}",
  journal = {\apj},
  year = {1992},
  volume = {395},
  pages = {540--552},
  doi = {10.1086/171674}
}

@article{chevalier2005,
  author = {{Chevalier}, R. A.},
  title = "{Young Core-collapse Supernova Remnants and Their Supernovae}",
  journal = {\apj},
  year = {2005},
  volume = {619},
  pages = {839--855},
  doi = {10.1086/426584}
}

@article{liu2021,
  author = {{Liu}, L.-D. and {Gao}, H. and {Wang}, X.-F. and {Yang}, S.},
  title = "{Magnetar-driven Shock Breakout Revisited and Implications for Double-peaked Type I Superluminous Supernovae}",
  journal = {\apj},
  year = {2021},
  volume = {911},
  pages = {142},
  doi = {10.3847/1538-4357/abf042}
}

@article{matzner1999,
  author = {{Matzner}, C. D. and {McKee}, C. F.},
  title = "{The Expulsion of Stellar Envelopes in Core-collapse Supernovae}",
  journal = {\apj},
  year = {1999},
  volume = {510},
  pages = {379--403},
  doi = {10.1086/306571}
}

@article{chevalier1989,
  author = {{Chevalier}, R. A. and {Soker}, N.},
  title = "{Asymmetric Envelope Expansion of Supernova 1987A}",
  journal = {\apj},
  year = {1989},
  volume = {341},
  pages = {867--882},
  doi = {10.1086/167545}
}

@article{Liu2025,
  author = {{Liu}, L.-D. and {Zhang}, Y.-H. and {Yu}, Y.-W. and {Du}, Z.-X. and {Li}, J.-Y. and {Wu}, G.-L. and {Dai}, Z.-G.},
  title = "{TransFit: An Efficient Framework for Transient Light-curve Fitting with Time-dependent Radiative Diffusion}",
  journal = {\apj},
  year = {2025},
  volume = {992},
  pages = {20},
  doi = {10.3847/1538-4357/adfed6}
}

@article{bellm2018,
  author = {{Bellm}, E. C. and {Kulkarni}, S. R. and {Graham}, M. J. and others},
  title = "{The Zwicky Transient Facility: System Overview, Performance, and First Results}",
  journal = {\pasp},
  year = {2019},
  volume = {131},
  pages = {018002},
  doi = {10.1088/1538-3873/aaecbe}
}

@article{ivezic2019,
  author = {{Ivezi{\'c}}, {\v Z}. and {Kahn}, S. M. and {Tyson}, J. A. and others},
  title = "{LSST: From Science Drivers to Reference Design and Anticipated Data Products}",
  journal = {\apj},
  year = {2019},
  volume = {873},
  pages = {111},
  doi = {10.3847/1538-4357/ab042c}
}

@article{gal2019,
  author = {{Gal-Yam}, A.},
  title = "{The Most Luminous Supernovae}",
  journal = {\araa},
  year = {2019},
  volume = {57},
  pages = {305--333},
  doi = {10.1146/annurev-astro-081817-051819}
}

@article{drout2014,
  author = {{Drout}, M. R. and {Chornock}, R. and {Soderberg}, A. M. and others},
  title = "{Rapidly Evolving and Luminous Transients from Pan-STARRS1}",
  journal = {\apj},
  year = {2014},
  volume = {794},
  pages = {23},
  doi = {10.1088/0004-637x/794/1/23}
}

@article{arnett1980,
  author = {{Arnett}, W. D.},
  title = "{Analytic Solutions for Light Curves of Supernovae of Type II}",
  journal = {\apj},
  year = {1980},
  volume = {237},
  pages = {541--549},
  doi = {10.1086/157898}
}

@article{arnett1982,
  author = {{Arnett}, W. D.},
  title = "{Type I Supernovae. I. Analytic Solutions for the Early Part of the Light Curve}",
  journal = {\apj},
  year = {1982},
  volume = {253},
  pages = {785--797},
  doi = {10.1086/159681}
}

@article{woosley2010,
  author = {{Woosley}, S. E.},
  title = "{Bright Supernovae from Magnetar Birth}",
  journal = {\apjl},
  year = {2010},
  volume = {719},
  pages = {L204--L207},
  doi = {10.1088/2041-8205/719/2/l204}
}

@article{chatzopoulos2012,
  author = {{Chatzopoulos}, E. and {Wheeler}, J. C. and {Vinko}, J.},
  title = "{Generalized Semi-analytical Models of Supernova Light Curves}",
  journal = {\apj},
  year = {2012},
  volume = {746},
  pages = {121},
  doi = {10.1088/0004-637x/746/2/121}
}

@article{nicholl2017,
  author = {{Nicholl}, M. and {Guillochon}, J. and {Berger}, E.},
  title = "{The Magnetar Model for Type I Superluminous Supernovae. I. Bayesian Analysis of the Full Multicolor Light-curve Sample with MOSFiT}",
  journal = {\apj},
  year = {2017},
  volume = {850},
  pages = {55},
  doi = {10.3847/1538-4357/aa9334}
}

@article{chen2016,
  author = {{Chen}, K.-J. and {Woosley}, S. E. and {Sukhbold}, T.},
  title = "{Magnetar-powered Supernovae in Two Dimensions. I. Superluminous Supernovae}",
  journal = {\apj},
  year = {2016},
  volume = {832},
  pages = {73},
  doi = {10.3847/0004-637x/832/1/73}
}

@article{morozova2015,
  author = {{Morozova}, V. and {Piro}, A. L. and {Renzo}, M. and others},
  title = "{Light Curves of Core-collapse Supernovae with Substantial Mass Loss Using the New Open-source SuperNova Explosion Code (SNEC)}",
  journal = {\apj},
  year = {2015},
  volume = {814},
  pages = {63},
  doi = {10.1088/0004-637x/814/1/63}
}

@article{blinnikov2006,
  author = {{Blinnikov}, S. I. and {R{\"o}pke}, F. K. and {Sorokina}, E. I. and others},
  title = "{Theoretical Light Curves for Deflagration Models of Type Ia Supernova}",
  journal = {\aap},
  year = {2006},
  volume = {453},
  pages = {229--240},
  doi = {10.1051/0004-6361:20054594}
}

@article{dessart2012,
  author = {{Dessart}, L. and {Hillier}, D. J. and {Waldman}, R. and {Livne}, E. and {Blondin}, S.},
  title = "{Superluminous Supernovae: $^{56}$Ni Power versus Magnetar Radiation}",
  journal = {\mnras},
  year = {2012},
  volume = {426},
  pages = {L76--L80},
  doi = {10.1111/j.1745-3933.2012.01329.x}
}

@article{pinto2000a,
  author = {{Pinto}, P. A. and {Eastman}, R. G.},
  title = "{The Physics of Type Ia Supernova Light Curves. I. Analytic Results and Time Dependence}",
  journal = {\apj},
  year = {2000},
  volume = {530},
  pages = {744--756},
  doi = {10.1086/308376}
}

@article{pinto2000b,
  author = {{Pinto}, P. A. and {Eastman}, R. G.},
  title = "{The Physics of Type Ia Supernova Light Curves. II. Opacity and Diffusion}",
  journal = {\apj},
  year = {2000},
  volume = {530},
  pages = {757--776},
  doi = {10.1086/308380}
}

@article{zhang2026,
  author = {{Zhang}, Y.-H. and {Liu}, L.-D. and {Du}, Z.-X. and {Wu}, G.-L. and {Li}, J.-Y. and {Yu}, Y.-W.},
  title = "{TransFit-CSM: A Fast, Physically Consistent Framework for Interaction-powered Transients}",
  journal = {\apj},
  year = {2026},
  volume = {999},
  pages = {186},
  doi = {10.3847/1538-4357/ae434a}
}

@article{zhang2022,
  author = {{Zhang}, Z.-D. and {Yu}, Y.-W. and {Liu}, L.-D.},
  title = "{The Effects of a Magnetar Engine on the Gamma-Ray Burst-associated Supernovae: Application to Double-peaked SN 2006aj}",
  journal = {\apj},
  year = {2022},
  volume = {936},
  pages = {54},
  doi = {10.3847/1538-4357/ac8548}
}

@article{li2016,
  author = {{Li}, S.-Z. and {Yu}, Y.-W.},
  title = "{Shock Breakout Driven by the Remnant of a Neutron Star Binary Merger: An X-Ray Precursor of Mergernova Emission}",
  journal = {\apj},
  year = {2016},
  volume = {819},
  pages = {120},
  doi = {10.3847/0004-637x/819/2/120}
}

@article{galama1998,
  author = {{Galama}, T. J. and {Vreeswijk}, P. M. and {van Paradijs}, J. and others},
  title = "{An Unusual Supernova in the Error Box of the {$\gamma$}-Ray Burst of 25 April 1998}",
  journal = {\nat},
  year = {1998},
  volume = {395},
  pages = {670--672},
  doi = {10.1038/27150}
}

@article{hjorth2003,
  author = {{Hjorth}, J. and {Sollerman}, J. and {M{\o}ller}, P. and others},
  title = "{A Very Energetic Supernova Associated with the {$\gamma$}-Ray Burst of 29 March 2003}",
  journal = {\nat},
  year = {2003},
  volume = {423},
  pages = {847--850},
  doi = {10.1038/nature01750}
}

@article{stanek2003,
  author = {{Stanek}, K. Z. and {Matheson}, T. and {Garnavich}, P. M. and others},
  title = "{Spectroscopic Discovery of the Supernova 2003dh Associated with GRB 030329}",
  journal = {\apjl},
  year = {2003},
  volume = {591},
  pages = {L17--L20},
  doi = {10.1086/376976}
}

@article{woosley2006,
  author = {{Woosley}, S. E. and {Bloom}, J. S.},
  title = "{The Supernova--Gamma-Ray Burst Connection}",
  journal = {\araa},
  year = {2006},
  volume = {44},
  pages = {507--556},
  doi = {10.1146/annurev.astro.43.072103.150558}
}

@article{yu2015,
  author = {{Yu}, Y.-W. and {Li}, S.-Z. and {Dai}, Z.-G.},
  title = "{Rapidly Evolving and Luminous Transients Driven by Newly Born Neutron Stars}",
  journal = {\apjl},
  year = {2015},
  volume = {806},
  pages = {L6},
  doi = {10.1088/2041-8205/806/1/L6}
}

@article{yu2017,
  author = {{Yu}, Y.-W. and {Zhu}, J.-P. and {Li}, S.-Z. and {L\"u}, H.-J. and {Zou}, Y.-C.},
  title = "{A Statistical Study of Superluminous Supernovae Using the Magnetar Engine Model and Implications for Their Connection with Gamma-Ray Bursts and Hypernovae}",
  journal = {\apj},
  year = {2017},
  volume = {840},
  pages = {12},
  doi = {10.3847/1538-4357/aa6c27}
}

@article{liu2022,
  author = {{Liu}, J.-F. and {Zhu}, J.-P. and {Liu}, L.-D. and {Yu}, Y.-W. and {Zhang}, B.},
  title = "{Magnetar Engines in Fast Blue Optical Transients and Their Connections with SLSNe, SNe Ic-BL, and lGRBs}",
  journal = {\apjl},
  year = {2022},
  volume = {935},
  pages = {L34},
  doi = {10.3847/2041-8213/ac86d2}
}

@article{yu2013,
  author = {{Yu}, Y.-W. and {Zhang}, B. and {Gao}, H.},
  title = "{Bright `Merger-nova' from the Remnant of a Neutron Star Binary Merger: A Signature of a Newly Born, Massive, Millisecond Magnetar}",
  journal = {\apjl},
  year = {2013},
  volume = {776},
  pages = {L40},
  doi = {10.1088/2041-8205/776/2/L40}
}

@article{pursiainen2018,
  author = {{Pursiainen}, M. and {Childress}, M. and {Smith}, M. and {Prajs}, S. and {Sullivan}, M. and {Davis}, T. M. and {M{\"o}ller}, A. and {Yuan}, F. and {Asorey}, J. and {Castander}, F. J. and others},
  title = "{Rapidly Evolving Transients in the Dark Energy Survey}",
  journal = {\mnras},
  year = {2018},
  volume = {481},
  pages = {894--917},
  doi = {10.1093/mnras/sty2309}
}

@article{margutti2019,
  author = {{Margutti}, R. and {Metzger}, B. D. and {Chornock}, R. and {Vurm}, I. and {Roth}, N. and {Grefenstette}, B. W. and {Savchenko}, V. and {Margalit}, B. and {Milisavljevic}, D. and {Alexander}, K. D. and others},
  title = "{An Embedded X-Ray Source Shines through the Aspherical AT 2018cow: Revealing the Inner Workings of the Most Luminous Fast-evolving Optical Transients}",
  journal = {\apj},
  year = {2019},
  volume = {872},
  pages = {18},
  doi = {10.3847/1538-4357/aafa01}
}

@article{ho2019,
  author = {{Ho}, A. Y. Q. and {Phinney}, E. S. and {Ravi}, V. and {Kulkarni}, S. R. and {Petitpas}, G. and {Emonts}, B. and {Bauer}, F. E. and {Cenko}, S. B. and {Corsi}, A. and {Dong}, D. and others},
  title = "{AT2018cow: A Luminous Millimeter Transient}",
  journal = {\apj},
  year = {2019},
  volume = {871},
  pages = {73},
  doi = {10.3847/1538-4357/aaf473}
}

@article{sun2025,
  author = {{Sun}, H. and {Li}, W.-X. and {Liu}, L.-D. and {Gao}, H. and {Wang}, X.-F. and {Yuan}, W. and {Zhang}, B. and {Filippenko}, A. V. and {Xu}, D. and {An}, T. and others},
  title = "{A Fast X-Ray Transient from a Weak Relativistic Jet Associated with a Type Ic-BL Supernova}",
  journal = {Nature Astronomy},
  year = {2025},
  volume = {9},
  pages = {1073--1085},
  doi = {10.1038/s41550-025-02571-1}
}

@article{li2025,
  author = {{Li}, W.-X. and {Zhu}, Z.-P. and {Zou}, X.-Z. and {Geng}, J.-J. and {Liu}, L.-D. and {Wang}, Y.-H. and {Li}, R.-Z. and {Xu}, D. and {Sun}, H. and {Wang}, X.-F. and {Yu}, Y.-W. and {Zhang}, B. and others},
  title = "{An Extremely Soft and Weak Fast X-Ray Transient Associated with a Luminous Supernova}",
  journal = {arXiv e-prints},
  year = {2025},
  eid = {arXiv:2504.17034},
  pages = {arXiv:2504.17034},
  archivePrefix = {arXiv},
  eprint = {2504.17034},
  primaryClass = {astro-ph.HE}
}

@article{zhu2026,
  author = {{Zhu}, J.-P. and {Zhang}, B.},
  title = "{Magnetar Engines in Broad-lined Type Ic Supernovae and a Unified Picture for Magnetar-powered Stripped-envelope Supernovae}",
  journal = {arXiv e-prints},
  year = {2026},
  eid = {arXiv:2604.21759},
  pages = {arXiv:2604.21759},
  archivePrefix = {arXiv},
  eprint = {2604.21759},
  primaryClass = {astro-ph.HE},
  doi = {10.48550/arXiv.2604.21759}
}

@article{Kashiyama2016,
  author        = {Kashiyama, Kazumi and Murase, Kohta and Bartos, Imre and Kiuchi, Kenta and Margutti, Raffaella},
  title         = {Multi-Messenger Tests for Fast-Spinning Newborn Pulsars Embedded in Stripped-Envelope Supernovae},
  journal       = {\apj},
  volume        = {818},
  number        = {1},
  pages         = {94},
  year          = {2016},
  doi           = {10.3847/0004-637X/818/1/94},
  archivePrefix = {arXiv},
  eprint        = {1508.04393},
  primaryClass  = {astro-ph.HE}
}
\bibliographystyle{aasjournalv7}

\end{document}